\documentclass[notitlepage,showpacs,aps,prl,eprint,superscriptaddress,nofootinbib]{revtex4-1}

\usepackage{epsfig,verbatim}
\usepackage{subfigure}
\usepackage{amsmath, amssymb, graphics}
\usepackage{color}
\usepackage{slashed}            
\usepackage{bbm}                
\usepackage{units}              
\usepackage{xspace}             
\usepackage{enumerate}          
\usepackage{soul}

\newcommand{\mathsym}[1]{{}}

\renewcommand\({\left(}
\renewcommand\){\right)}
\renewcommand\[{\left[}
\renewcommand\]{\right]}

\renewcommand{\Re}{\operatorname{Re}}

\newcommand{\la}{\langle}
\newcommand{\ra}{\rangle}

\newcommand{\eref}[1]{\eqref{#1}}

\newcommand{\mode}[1]{\ensuremath{\mathcal{#1}}\xspace}

\newcommand\eps{\epsilon}

\newcommand{\ba}{\begin{eqnarray}}
\newcommand{\ea}{\end{eqnarray}}
\newcommand{\be}{\begin{equation}}
\newcommand{\ee}{\end{equation}}

\newcommand{\nn}{\nonumber}

\newcommand{\codename}[1]{\textsc{#1}\xspace}
\newcommand{\class}{\codename{CLASS}}
\newcommand{\montepython}{\codename{Monte Python}}
\newcommand{\camb}{\codename{CAMB}}
\newcommand{\cosmomc}{\codename{CosmoMC}}
\newcommand{\lcdm}{$\Lambda$CDM\xspace}

\begin{document}



\title{Inflation with moderately sharp features in
  the speed of sound:\\ GSR and in-in formalism for power spectrum and
  bispectrum}

\author{Ana Ach\'ucarro}
\email[]{achucar@lorentz.leidenuniv.nl}
\affiliation{Instituut-Lorentz for Theoretical Physics, Universiteit Leiden, 2333 CA Leiden, The Netherlands}
\affiliation{Department of Theoretical Physics, University of the Basque Country, 48080 Bilbao, Spain}
\author{Vicente Atal}
\email[]{atal@lorentz.leidenuniv.nl}
\affiliation{Instituut-Lorentz for Theoretical Physics, Universiteit Leiden, 2333 CA Leiden, The Netherlands}
\author{Bin Hu}
\email[]{hu@lorentz.leidenuniv.nl}
\affiliation{Instituut-Lorentz for Theoretical Physics, Universiteit Leiden, 2333 CA Leiden, The Netherlands}
\author{Pablo Ortiz}
\email[]{ortiz@lorentz.leidenuniv.nl}
\affiliation{Instituut-Lorentz for Theoretical Physics, Universiteit Leiden, 2333 CA Leiden, The Netherlands}
\affiliation{Nikhef, Science Park 105, 1098 XG Amsterdam, The Netherlands}
\author{Jes\'us Torrado}
\email[]{torradocacho@lorentz.leidenuniv.nl}
\affiliation{Instituut-Lorentz for Theoretical Physics, Universiteit Leiden, 2333 CA Leiden, The Netherlands}





\date{\today}

\begin{abstract}

We continue the study of mild transient reductions in the
speed of sound of the adiabatic mode during inflation, of their effect on
the primordial power spectrum and bispectrum, and of their
detectability in the Cosmic Microwave Background (CMB).
We focus on the regime of {\it moderately sharp} mild reductions in the
speed of sound during uninterrupted slow-roll inflation, a
theoretically well motivated and self-consistent regime that admits an
effective single-field description. The signatures on the power spectrum and bispectrum were previously
computed using a slow-roll Fourier transform (SRFT) approximation, and
here we compare it with generalized slow-roll (GSR) and in-in methods, for which
we derive new formulas that account for moderately sharp features.  The
agreement between them is excellent, and also with the power spectrum
obtained from the numerical solution to the equation of motion.
We show that, in this regime, the SRFT approximation correctly captures
with simplicity the effect of higher derivatives of the speed of sound
in the mode equation, 
and makes
manifest the correlations between power spectrum and bispectrum
features.  In a previous paper we reported hints of these correlations
in the Planck data and here we perform several consistency checks
and further analyses of the best fits, such as polarization and local
significance at different angular scales. For the data analysis, we
show the excellent agreement between the CLASS and CAMB Boltzmann
codes. Our results confirm that the theoretical framework is
consistent, and they suggest that the predicted correlations are robust enough to
be searched for in CMB and Large Scale Structure (LSS) surveys.

\end{abstract}



\maketitle





\section{I. Introduction}

The paradigm of inflation as the explanation for the origin of
cosmic structures has entered a decisive new phase. The latest data releases by the Planck \cite{Ade:2013ktc} and WMAP \cite{Bennett:2012zja} collaborations point towards models of
inflation that produce a slightly red-tilted primordial power spectrum
and a negligible amount of \emph{scale-independent} bispectra, as
predicted \cite{Mukhanov:1981xt,Acquaviva:2002ud,Maldacena:2002vr} by
the simplest models of cosmological inflation,\footnote{These are
  slow-roll inflation models involving a single neutral scalar field
  with a canonical kinetic term and in the Bunch-Davies vacuum.} but
with a mild deficit of power on large scales. There are also mild
hints of \emph{scale-dependent} features in the CMB power spectrum \cite{Bennett:2012zja,Ade:2013uln} and in the
primordial
bispectrum \cite{Ade:2013ydc}. Besides this, the discovery of B-mode polarization by BICEP2
\cite{Ade:2014xna}, if it is confirmed to be result of primordial
tensor modes, would have striking implications and put inflation on a
much firmer footing. A large tensor-to-scalar ratio of $r\sim{\cal O}(0.1)$
suggests -- again, in the context of canonical models -- a high scale
of inflation around $10^{16}\,\text{GeV}$, a Hubble parameter $H \sim 10^{14}\,\text{GeV}$ during inflation and a large, transplanckian excursion in field
space for the inflaton \cite{Lyth:1996im}.

According to \cite{Smith:2014kka}, there is currently a ``very significant tension'' (around $0.1\%$ unlikely)  
between the Planck temperature ($r<0.11~95\%$c.l.) and BICEP2 polarization ($r=0.2^{+0.05}_{-0.07}$) 
results. The model-independent cubic spline reconstruction \cite{Hu:2014aua} result shows that the vanishing 
scalar index running (${\rm d n_s}/{\rm d}\ln k$) model is strongly disfavored at more than $3\sigma$ confidence 
level on the scales $k=0.0002$ Mpc$^{-1}$. Recently, several fundamental/phenomenological models 
with features in the primordial spectra, such as sharp transition in the slow-roll parameters \cite{Miranda:2014wga}, false vacuum decay \cite{Hazra:2014aea,Bousso:2014jca}, initial fast roll \cite{Hazra:2014aea}, a non-Bunch-Davies initial state \cite{Ashoorioon:2014nta}, or a bounce before inflation \cite{Xia:2014tda}, among others, were proposed to explain the observed power deficit on large angular scales by Planck experiments. Alternatively, the tension could be resolved with new data releases. 

Another consequence of the BICEP2 results is that a large
tensor-to-scalar ratio seems to indicate a high energy scale of
inflation around the GUT scale. If confirmed, one would need to find
a successful UV embedding of the theory, and also deal with the
problem of mass hierarchies in the presence of multiple degrees of
freedom. This is challenging, but not impossible, and it seems that the
energy range available could in principle host the inflaton and the
possible additional UV degrees of freedom, while preserving a
manageable mass hierarchy for which an effective single field theory
is still possible. The BICEP2 results also suggest that the
inflaton field underwent a super-planckian excursion, which makes the
theory very sensitive to higher dimensional operators. While we expect
a (mildly broken) symmetry protecting the overall flatness of the
potential, this also leaves room for the presence of transient
phenomena happening along the inflationary trajectory.

Among other phenomena, transient variations in the speed sound of the adiabatic mode may occur in the presence of additional degrees of freedom during inflation. For instance, when an additional heavy field can be consistently integrated out \cite{Shiu:2011qw,Cespedes:2012hu,Achucarro:2012sm,Achucarro:2012yr,Burgess:2012dz,Castillo:2013sfa} (see also \cite{Ashoorioon:2008qr}), inflation is described by an effective single-field theory \cite{Burgess:2003zw,Cheung:2007st,Weinberg:2008hq,Shiu:2011qw,Achucarro:2012sm,Achucarro:2012yr} with a variable speed of sound. In particular, changes in the speed of sound result from derivative couplings\footnote{Or equivalently, turns in field space.} \cite{Tolley:2009fg,Achucarro:2010da,Baumann:2011su,Cespedes:2012hu,Achucarro:2012sm,Achucarro:2012yr,Pi:2012gf,Chen:2012ge,Gao:2012uq}. Transient variations in the speed of sound will produce \emph{correlated} features in the correlation functions of the adiabatic curvature perturbation \cite{Cheung:2007st,Cai:2009hw,Nakashima:2010sa,Park:2012rh,Achucarro:2012fd,Saito:2013aqa,Bartolo:2013exa,Cai:2013gma,Gong:2014spa}. They are worth taking into account since we expect them to be very good model selectors.

The detection of transients poses some interesting
challenges. The effects of a feature in the potential or a localized change
in the speed of sound depend on its {\it location} (in time or
e-folds), its {\it amplitude} and the {\it sharpness} (or inverse
duration). If transients are too sharp, they can excite higher
frequency modes that make the single-field interpretation inconsistent
(see, for example, \cite{Cespedes:2012hu,Shiu:2011qw,Konieczka:2014zja}).
Notably, the best fit found so far in the data for a step
feature in the potential
\cite{Ade:2013uln,Benetti:2013cja,Miranda:2013wxa} falls outside the
weakly coupled regime that is implicitly required for its
interpretation as a step in the single field potential \cite{Adshead:2014sga,Cannone:2014qna}. On the other hand, if the
features are too broad, their signature usually becomes degenerate
with cosmological parameters, making their presence difficult to
discern.  There is an interesting intermediate regime where the
features are mild (small amplitude) and moderately sharp, which makes them potentially
detectable in the CMB/LSS data, and also they remain under good
theoretical control. This regime is particularly important if the inflaton field
excursion is large and can reveal features in the inflationary
potential and the presence of other degrees of freedom. At the same
time, if slow-roll is the result of a (mildly broken) symmetry that
protects the background in the UV completion, the same symmetry might
presumably preclude very sharp transients.

In this paper we study \emph{mild and moderately sharp} features in the speed
of sound of the adiabatic mode, that we define to be those for which
the effects coming from a varying speed of sound are small enough to
be treated at linear order, but large enough to dominate over the
slow-roll corrections. This carries an implicit assumption of
uninterrupted slow-roll.\footnote{In the particular case of reductions in the speed of sound coming from turns
along the inflationary trajectory, this has been shown to be a consistent scenario.} We will show that this regime
ensures the validity of the effective single-field theory, even though
our analysis is blind to the underlying
inflationary model.

In order to compare any model with data, it is important to develop fast and accurate techniques to compute the relevant
observables of the theory, in this case, correlations functions of the
adiabatic curvature perturbation. The calculation of correlation functions is often rather complicated
and the use of approximate methods is needed. The study of transients
often involves deviations from slow-roll and may be analyzed in the
generalized slow-roll (GSR) formalism
\cite{Stewart:2001cd,Gong:2001he,Choe:2004zg,Dvorkin:2009ne,Adshead:2011bw,Miranda:2012rm,Bartolo:2013exa,Adshead:2013zfa,Gong:2014spa}. This
approach is based on solving the equations of motion iteratively using
Green's functions. Although this formalism can cope with  more
general situations with both slow-roll and speed of sound features,
one usually needs to impose extra hierarchies between the different
parameters to obtain simple analytic solutions.

A notable exception that is theoretically well understood is a
transient, mild, and moderately sharp reduction in the speed of sound
such as would be found in effectively single-field models with
uninterrupted slow-roll inflation, obtained by integrating out much
heavier fields with derivative couplings that become transiently
relevant. In this regime, an alternative approach is possible, that
makes the correlation between power spectrum and bispectrum manifest
\cite{Achucarro:2012fd}. The change in the power spectrum is simply
given by the Fourier transform of the reduction in the speed of sound,
and the \emph{complete} bispectrum can be calculated to leading order
in slow-roll as a function of the power spectrum. Hence we name this
approximation Slow-Roll Fourier Transform (SRFT). One of the aims of this
paper is to compare the GSR and SRFT approaches. In order to do this, we develop
simple expressions within the GSR approach and the in-in formalism for
computing the changes in the power spectrum and bispectrum due to
moderately sharp features in the speed of sound. These are new and extend the usual GSR expressions for very sharp features.

The other aim of this paper is to further scrutinize and validate the results of our previous work \cite{Achucarro:2013cva}, where we searched for moderately sharp features in the Planck CMB data. We reported several fits to
the CMB power spectrum and gave the predicted, correlated, oscillatory
signals for the primordial bispectrum. The functional form of the
speed of sound was inspired by soft turns along a multi-field
inflationary trajectory with a large hierarchy of masses, a situation
that is consistently described by an effective single-field theory
\cite{Achucarro:2010jv,Achucarro:2010da,Cespedes:2012hu,Achucarro:2012sm,Gao:2012uq,Gao:2013ota}.

In the first part of this paper we study the intermediate regime of moderately sharp features in the speed of sound during uninterrupted slow-roll, in which both the SRFT and GSR approaches can give accurate results. More precisely:

\begin{itemize}

\item In \S IIA we review the SRFT results for the power spectrum and bispectrum, and in \S IIB we develop a simple formula within the GSR formalism that reduces to the SRFT result for nearly all scales and is valid for arbitrary functional forms of the speed of sound within the regime we study.

\item In \S IIC, by comparing both results with a numerical solution for the power spectrum, we show that the SRFT method correctly captures the effect of \emph{all} the terms in the equation of motion in a very simple way, while the GSR method requires the inclusion of higher derivatives of the speed of sound to match the numerical result. Nevertheless, there is excellent agreement between both results with the numerical solution.

\item Then we turn to the bispectrum. In \S IID we compute the features in the bispectrum using the in-in formalism, and we take into account the effect of additional operators with respect to previous results \cite{Bartolo:2013exa}. We show that, for transient reductions of the speed of sound, the contributions arising from the operators proportional to the amount of reduction and to the rate of change are of the same order, \emph{independently of the sharpness} of the feature. In addition, because we study the not-so-sharp regime, we compute the linear correction to the approximation that other quantities do not vary during the time when the feature happens.

\item In \S IIE we compare the bispectra obtained with the SRFT approach and with the moderately sharp approximation, finding remarkable agreement for several functional forms of the speed of sound.
\end{itemize}

In the second part of this paper we perform a number of additional consistency checks regarding the theoretical framework and the statistical analysis carried out in a previous paper \cite{Achucarro:2013cva}. In particular:

\begin{itemize}

\item In \S IIIA we explain the choice of parameter space used for our statistical search of transient reductions of the speed of sound in the Planck data, which was designed to be theoretically consistent. In \S IIIB we check that adiabatic and unitary regimes are respected, and therefore the fits found in the data can be consistently interpreted as transient reductions in the speed of sound.

\item In \S IIIC we analyze the implications of the BICEP2 results for the consistency of an effective single-field description of inflation. We conclude that, even with a inflationary scale at the level of the GUT scale, a single-field description may be possible, and we argue that moderately sharp reductions of the speed of sound are completely consistent with an adiabatic evolution, i.e. an effective single-field regime.

\item In \S IIID we review the main results of our previous work \cite{Achucarro:2013cva} and make an independent consistency check using two different Boltzmann codes and MCMC samplers, namely \class\unskip$+$\montepython versus \camb\unskip$+$\cosmomc, finding great agreement. We explicitly give the (small) degeneracy of the cosmological parameters with the parameters of our model. Last, we also show the polarization spectra and the local improvement of our fits to the CMB power spectrum as a function of the angular scale.

\end{itemize}

Finally, we leave \S IV for conclusions and outlook.


\section{II. Moderately sharp variations in the speed of sound: primordial power spectrum and bispectrum}


In the framework of the effective field theory (EFT) of inflation \cite{Cheung:2007st} one can write the effective action for the Goldstone boson of time diffeomorphisms $\pi(t,{\bf x})$, directly related to the adiabatic curvature perturbation ${\cal R}(t,{\bf x})$ via the linear relation\footnote{In this work, we do not need to consider non-linear correction terms, since we are in a slow-roll regime. For further details on this, see \cite{Maldacena:2002vr}.} ${\cal R}=-H\pi$. Let us focus on a slow-roll regime and write the quadratic and cubic actions for $\pi$:
\ba
S_2&=&\int d^4x\, a^3 M_\text{Pl}^2\epsilon H^2\left\{\frac{\dot\pi^2}{c_s^2}-\frac{1}{a^2}\(\nabla\pi\)^2\right\}\ ,\label{s2}\\\nn \\
S_3&=&\int d^4x\, a^3M_\text{Pl}^2\epsilon H^2\left\{-2Hsc_s^{-2}\pi\dot\pi^2-\(1-c_s^{-2}\)\dot\pi\[\dot\pi^2-\frac{1}{a^2}\(\nabla\pi\)^2\]\right\}\ ,\label{s3}
\ea
where $\eps=-\dot H/H^2$ and we are neglecting higher order slow-roll corrections, as well as higher order terms in $u$ and $s$, defined as:
\be
u\equiv 1-c_s^{-2}\quad,\quad s\equiv\frac{\dot c_s}{c_sH}\ .
\ee
In this section we compare the different approaches to evaluating the power spectrum and bispectrum of the adiabatic curvature perturbation from \eref{s2} and \eref{s3} with a variable speed of sound, and show the excellent agreement between them.

The Slow-Roll Fourier Transform (SRFT) approach, developed in \cite{Achucarro:2012fd}, is briefly reviewed in \S IIA. The advantage of this method is that one obtains very simple analytic formulas for both the power spectrum and bispectrum computed from \eref{s2} and \eref{s3}. More importantly, correlations between features in the power spectrum and bispectrum show up explicitly. In \S II B we review the Generalized Slow Roll (GSR) formalism \cite{Stewart:2001cd,Gong:2001he,Adshead:2011jq,Hu:2011vr,Park:2012rh,Bartolo:2013exa,Adshead:2013zfa,Gong:2014spa} and compute the power spectrum from the quadratic action \eref{s2} in the moderately sharp approximation. This method applies to more general situations where slow-roll is not necessarily preserved, but it requires solving iteratively the equations of motion, which include higher derivatives of the speed of sound. The GSR formalism gives very simple expressions in the case of very sharp features and has been used to calculate the effect of steps in the potential and in the speed of sound (see for example \cite{Bartolo:2013exa,Miranda:2012rm}).

In \S II C we compare both methods with the power spectrum obtained from the numerical solution to the mode equations. We show that the SRFT method correctly captures the effect of higher derivative terms of the speed of sound in a very simple way, while the GSR method requires the inclusion of all terms in the equations of motion to match the numerical result at all scales (especially at the largest scales).

Then we turn to the bispectrum. In \S II D we compute the bispectrum from the cubic action \eref{s3} using an approximation for sharp features as in \cite{Bartolo:2013exa}, but including the next order correction and additional operators. Last, in \S II E we check that the agreement with the SRFT result \cite{Achucarro:2012fd} is excellent. An important point we show is that the contributions to the bispectrum arising from the terms proportional to $(1-c_s^{-2})$ and $s$ in \eref{s3} are of the same order, {\it independently of the sharpness of the feature}. We also eliminate the small discrepancy found in \cite{Bartolo:2013exa} between their bispectrum and the one obtained with GSR \cite{Adshead:2011jq} for step features in the scalar potential, due to a missing term in the bispectrum.


\subsection{A. Power spectrum and bispectrum with the SRFT method}


In this formalism \cite{Achucarro:2012fd} we assume an uninterrupted slow-roll regime, which is perfectly consistent with turns along the inflationary trajectory. In order to calculate the power spectrum, we separate the quadratic action \eref{s2} in a free part and a small perturbation:
\be
S_2=\int d^4x\, a^3 M_\text{Pl}^2\epsilon H^2\left\{\dot\pi^2-\frac{1}{a^2}\(\nabla\pi\)^2\right\}-\int d^4x\, a^3 M_\text{Pl}^2\epsilon H^2\bigg\{\dot\pi^2\(1-c_s^{-2}\)\bigg\}\ ,\label{s2sep}
\ee
Then, using the in-in formalism \cite{Keldysh:1964ud,Weinberg:2005vy}, the change in the power spectrum due to a small transient reduction in the speed of sound can be calculated to first order in $u\equiv1-c_s^{-2}$, and it is found to be \cite{Achucarro:2012fd}
\begin{equation}\label{eq:deltappfourier}
\frac{\Delta {\cal P_R}}{{\cal P}_{{\cal R},0}}(k)=k\int_{-\infty}^{0}d\tau\ u(\tau)\sin{(2k\tau)}\ ,
\end{equation}
where $k\equiv|{\bf k}|$, ${\cal P}_{{\cal R},0}=H^2/(8\pi^2\epsilon M_{\text{Pl}}^2)$ is the featureless power spectrum with $c_s=1$, and $\tau$ is the conformal time. We made the implicit assumption that the speed of sound approaches to one asymptotically, since we are perturbing around that value.\footnote{At the level of the power spectrum, the generalization to arbitrary initial and final values of the speed of sound $c_{s,0}$ is straightforward, provided they are sufficiently close to each other.} Here we see that the change in the power spectrum is simply given by the Fourier transform of the reduction in the speed of sound.
Notice that the result above is independent of the physical origin of such reduction.

For the three-point function, we take the cubic action \eref{s3}, written to first order in $u$ and $s$, which implies that we must have $|u|_\text{max},|s|_\text{max}\ll 1$. We also disregard the typical slow-roll contributions that one expects for a canonical featureless single-field regime \cite{Maldacena:2002vr}. Therefore, for the terms proportional to $u$ and $s$ to give the dominant contribution to the bispectrum, one must require that $u$ and/or $s$ are much larger than the slow-roll parameters, i.e. $\text{max}(u,s)\gg{\cal O}(\epsilon,\eta)$, as we will recall in \S III A. Using the in-in formalism, one finds \cite{Achucarro:2012fd}:
\ba\label{bispectrumana}
\Delta B_{\cal R}({\bf k}_1,{\bf k}_2,{\bf k}_3) = \frac{(2\pi)^4{\cal P}_{{\cal R},0}^2}{(k_1k_2k_3)^2} \left\{ -\frac{3}{2} \frac{k_1k_2}{k_3} \left[ \frac{1}{2k} \left( 1 + \frac{k_3}{2k} \right) \frac{\Delta{\cal P}_{\cal R}}{{\cal P}_{{\cal R},0}} - \frac{k_3}{4k^2} \frac{d}{d\log k} \left( \frac{\Delta{\cal P}_{\cal R}}{{\cal P}_{{\cal R},0}} \right) \right] + \text{\footnotesize [2 perm]}\hspace{2cm} \right.\\\nn\\
\left.  + \frac{1}{4} \frac{k_1^2+k_2^2+k_3^2}{k_1k_2k_3} \left[ \frac{1}{2k} \left( 4k^2 - k_1k_2 - k_2k_3 - k_3k_1 - \frac{k_1k_2k_3}{2k} \right) \frac{\Delta{\cal P}_{\cal R}}{{\cal P}_{{\cal R},0}}\hspace{2cm} \right. \right.
\nonumber\\\nn\\
 \left. \left. - \frac{k_1k_2+k_2k_3+k_3k_1}{2k} \frac{d}{d\log k} \left( \frac{\Delta{\cal P}_{\cal R}}{{\cal P}_{{\cal R},0}} \right) + \frac{k_1k_2k_3}{4k^2} \frac{d^2}{d\log k^2} \left( \frac{\Delta{\cal P}_{\cal R}}{{\cal P}_{{\cal R},0}} \right) \right] \right\}\Bigg|_{k=\tfrac{1}{2}\sum_ik_i}\nonumber \, ,
\ea
where $k_i\equiv |{\bf k}_i|$, $k\equiv (k_1+k_2+k_3)/2$, and ${\Delta{\cal P}_{\cal R}}/{{\cal P}_{{\cal R},0}}$ and its derivatives are evaluated at $k$. From the result above it is clear how features in the power spectrum seed correlated features in the bispectrum. Note that in the squeezed limit $(k_1\to 0, k_2=k_3=k)$ one recovers the single-field consistency relation \cite{Maldacena:2002vr,Creminelli:2004yq}.

In the following sections, we compute the power spectrum and bispectrum using alternative methods and compare the results.


\subsection{B. Power spectrum in the GSR formalism}


One can calculate the power spectrum by solving iteratively the full equations of motion (first in \cite{Gong:2001he,Stewart:2001cd} and further developed in \cite{Choe:2004zg,Dvorkin:2009ne,Adshead:2011jq,Hu:2011vr,Adshead:2013zfa,Gong:2014spa}). The idea is to consider the Mukhanov-Sasaki equation of motion with a time-dependent speed of sound, namely:
\be
\frac{d^2v_{\bf k}(\tau)}{d\tau^2}+\(c_s^2k^2-\frac{1}{z}\frac{d^2z}{d\tau^2}\)v_{\bf k}(\tau)=0\ ,
\label{MSequation}
\ee
with $v=z\mathcal{R}$, $z^2=2a^2M_\text{Pl}^2\epsilon c_s^{-2}$ and
\be
\frac{1}{z}\frac{d^2z}{d\tau^2}=a^2H^2\Big[2+2\epsilon-3\tilde\eta-3s+2\epsilon(\epsilon-2\tilde\eta-s)+s(2\tilde\eta+2s-t)+\tilde\eta\tilde\xi\,\Big]\ ,
\label{MSmassterm}
\ee
where we have used the following relations:
\be
\epsilon=-\frac{\dot H}{H^2}\quad ,\quad \tilde\eta=\epsilon-\frac{\dot\epsilon}{2H\epsilon}\quad ,\quad s=\frac{\dot c_s}{H c_s}\quad ,\quad t=\frac{\ddot c_s}{H\dot c_s}\quad ,\quad \tilde\xi=\epsilon+\tilde\eta-\frac{\dot{\tilde\eta}}{H\tilde\eta}\ ,
\label{MSdefs}
\ee
and here the dot denotes the derivative with respect to cosmic time. Defining a new time variable $d\tau_c=c_sd\tau$ and a rescaled field $y=\sqrt{2kc_s}v$, the above equation can be written in the form:
\be\label{MSvar}
\frac{d^2y}{d\tau_c^2}+\(k^2-\frac{2}{\tau_c^2}\)y=\frac{g\(\ln \tau_c\)}{\tau_c^2}y\ ,
\ee  
where
\be
g\equiv \frac{f''-3f'}{f} \quad\quad,\quad f=2\pi z c_s^{1/2} \tau_c\ ,
\ee
and $'$ denotes derivatives with respect to $\ln\tau_c$. Throughout this section (and only in this section), unless explicitly indicated, we will adopt the convention of positive conformal time ($\tau,\tau_c\geq0$) in order to facilitate comparison with \cite{Dvorkin:2009ne,Hu:2011vr}. Note that $g$ encodes all the information with respect to features in the background. In this sense, setting $g$ to zero represents solving the equation of motion for a perfect de Sitter universe, where the solution to the mode function is well known. Considering the r.h.s.\ of equation (\ref{MSvar}) as an external source, a solution to the mode function can be written in terms of the homogeneous solution. In doing so, we need to expand the mode  function in the r.h.s.\ as the homogeneous solution plus deviations and then solve iteratively. To first order, the contribution to the power spectrum is of the form \cite{Dvorkin:2009ne}:
\begin{equation}
\ln {\cal P}_{\cal R}=\ln {\cal P}_{{\cal R},0}+\int_{-\infty}^{\infty}d\ln\tau_c\,W\left(k\tau_c\right)G'\left(\tau_c\right),
\end{equation}
where the logarithmic derivative of the source function $G$ reads:
\be
G'=-2(\ln f)' +\frac{2}{3}(\ln f)''\ ,
\ee
and the window function $W$ and its logarithmic derivative (used below) are given by
\ba
W\left(x\right)&=&\frac{3\sin\left(2x\right)}{2x^3}-\frac{3\cos\left(2x\right)}{x^2}-\frac{3\sin\left(2x\right)}{2x}\ ,\\\nn\\
W'(x)&\equiv&\frac{dW(x)}{d\ln x}=\(-3+\frac{9}{x^2}\)\cos (2x)+\(\frac{15}{2x}-\frac{9}{2x^3}\)\sin (2x)\ .
\ea
If we consider moderately sharp features in the speed of sound, such that $\epsilon,\tilde\eta\ll s,t$, the leading contribution to the function $G'$ is the following:
\be\label{sourcefunction}
G'=-\frac{2}{3}s+\frac{2}{3}\left(\frac{aH\tau_c}{c_s}-1\right)^2+\frac{2}{3}\left(\frac{aH\tau_c}{c_s}-1\right)\left(4-s\right)+\frac{1}{3}\left(\frac{aH\tau_c}{c_s}\right)^2s\left(-3+2s-t\right)\ ,
\ee
where $t$ is defined in \eref{MSdefs}. Moreover, when $|s|\ll1$ but $t\gtrsim {\cal O}(1)$, the logarithmic derivative of $G$ is approximately given by:
\be
G'\simeq s-\frac{\dot s}{3H}
\ ,
\label{sourcesharp}
\ee
where we have used that $aH\tau_c/c_s\simeq 1+s$. This result agrees with the results of \cite{Hu:2011vr} in the mentioned limits. In this approximation, the leading contribution to the power spectrum is:
\begin{equation}
\ln {\cal P}_{\cal R}\simeq \ln {\cal P}_{{\cal R},0}+\int_{-\infty}^{\infty}d\ln\tau_c\,\left[W(k\tau_c)s\(\tau_c\)-\frac{1}{3}W\left(k\tau_c\right)\frac{d s}{d\ln\tau_c}\right] \ .
\end{equation}
Integrating by parts the term proportional to the derivative of $s$ we obtain:
\ba
\ln {\cal P}_{\cal R}\simeq \ln {\cal P}_{{\cal R},0}+\int_{-\infty}^{\infty}d\,\ln\tau_c\,\left[W\left(k\tau_c\right)+\frac{1}{3}W'\left(k\tau_c\right)\right]s\left(\tau_c\right)\nn
\\\nn\\
=\ln {\cal P}_{{\cal R},0}+\int_{-\infty}^{\infty}d\,\ln\tau_c\,\left[\frac{\sin (2k\tau_c)}{k\tau_c}-\cos (2k\tau_c)\right]s\left(\tau_c\right)\ .
\label{GsrInt2}
\ea
This is the result that we will compare in \S IIC with the SRFT result \eref{eq:deltappfourier}. Let us recall that the regime in which this expression has been derived is for moderately sharp reductions such that ${\cal O}(\epsilon,\eta)\ll s\ll 1$ and $t\gtrsim {\cal O}(1)$. We would like to point out that the $s$ term in the source function \eref{sourcesharp} provides the dominant contribution to the power spectrum on large scales. This can be seen by comparing $W$ and $W'$ in \eref{GsrInt2}, which carry the contribution of $s$ and $\dot s$, respectively. We will show in \S IIC that when including this term, the power spectrum at large scales matches the numerical solution considerably better (see figure \ref{fig:methods}).

In the following, we will: {\bf (i)} derive an analytic expression for the power spectrum \eref{GsrInt2} solely in terms of $c_s$ in order to connect with the SRFT approach. {\bf (ii)} Find an analytic approximation for arbitrary functional forms of the speed of sound in the moderately sharp regime specified above.

\medskip
{\bf (i)} For the first point, one can integrate by parts \eref{GsrInt2} in order to get a formula than only involves the speed of sound. Doing so, we obtain:
\be
\ln {\cal P}_{\cal R}=\ln {\cal P}_{{\cal R},0}-\int_{-\infty}^{\infty}d\,\ln\tau_c\,\left[2\cos (2k\tau_c)-\frac{\sin (2k\tau_c)}{k\tau_c}+2k\tau_c\sin (2k\tau_c)\right]\ln c_s\left(\tau_c\right)\ ,
\ee
 where we have used that $s\simeq d\ln c_s/d\ln \tau_c$ and that the asymptotic value of the speed of sound is one, otherwise the boundary term would not vanish. Therefore, the expression above is only valid for functional forms of the speed of sound that satisfy $c_s(\tau=0)=c_s(\tau=\infty)=1$. Let us restrict our attention to mild reductions of the speed of sound $|u|=|1-c_s^{-2}|\ll 1$, in which the SRFT approach is operative. In that case, for mild and moderately sharp reductions, the time $\tau_c$ is very well approximated by $\tau_c\simeq\tau$. Furthermore, the logarithmic term of the speed of sound can be expanded as follows:
\be
\ln c_s(\tau)\simeq \frac{1}{2}\(1-c_s^{-2}(\tau)\) + {\cal O}(u^2)\ .
\ee
Using the expansion above and the fact that $\ln  ({\cal P}_{{\cal R}}/{\cal P}_{{\cal R},0})=\ln(1+\Delta{\cal P}_{{\cal R}}/{\cal P}_{{\cal R},0})\simeq \Delta{\cal P}_{{\cal R}}/{\cal P}_{{\cal R},0}$, we can write:
\be
\frac{\Delta{\cal P}_{{\cal R}}}{{\cal P}_{{\cal R},0}}\simeq k\int_{-\infty}^{0}d\tau\,\(1- c_s^{-2}\)\left[\,\sin (2k\tau)+\frac{1}{k\tau}\cos (2k\tau)-\frac{1}{2k^2\tau^2}\sin (2k\tau)\right]{\simeq} \begin{cases} \frac{\Delta{\cal P}_{{\cal R}}}{{\cal P}_{{\cal R},0}}\Big|_\text{SRFT}+{\cal O}\[(k\tau)^2\]\ ,\ k\tau\ll 1 \\ \\ \frac{\Delta{\cal P}_{{\cal R}}}{{\cal P}_{{\cal R},0}}\Big|_\text{SRFT}+{\cal O}\[(k\tau)^{-1}\]\  ,\ k\tau\gg 1 \end{cases}
\label{GSRtoSRFT}
\ee
where we have already returned to negative conformal time. Notice that when $k\tau\ll1$ we retrieve the SRFT expression \eref{eq:deltappfourier} with a subleading correction ${\cal O}(k\tau)$ inside the integral, and that for $k\tau\gg1$ we also retrieve the SRFT result. The regime $k\tau\sim 1$ will generally involve large scales, where the change in the power spectrum is small, as can be seen in figure \ref{fig:methods}.

\medskip 
{\bf (ii)} In what follows we derive an analytic approximation to the power spectrum \eref{GsrInt2} for \emph{generic} forms of the speed of sound, provided they are moderately sharp, i.e. ${\cal O}(\epsilon,\eta)\ll s\ll 1$ and $t\gtrsim {\cal O}(1)$. As in {\bf (i)}, in this regime we can safely consider $\tau_c\simeq c_{s,0}\tau$. Let us drop the rest of assumptions made in point {\bf (i)}, which were only made to establish connection with the SRFT approach. We define the function $X(k\tau_c)\equiv -W'(k\tau_c)-3W(k\tau_c)$, which in general can be decomposed as follows:
\be
X(kc_{s,0}\tau)=p_c(kc_{s,0}\tau)\cos (2kc_{s,0}\tau)+p_s(kc_{s,0}\tau)\sin (2kc_{s,0}\tau)\ ,
\label{Xdef}
\ee
where $p_c$ and $p_s$ denote the polynomials multiplying the cosine and sine, respectively. 
Following \cite{Bartolo:2013exa}, we will parametrize $c_s^2$ in terms of the height $\sigma_*$ and the sharpness $\beta_s$ of the feature, and a function $F$ describing the shape of the variation of the speed of sound:
\be
c_s^2(\tau)=c_{s,0}^2\[1-\sigma_*F\(-\beta_s\ln\tfrac{\tau}{\tau_f}\)\]\ ,
\label{Fsharp}
\ee
where $\tau_f$ is the characteristic time of the feature and we take $\sigma_*\ll1$ to focus on small variations. The rate of change in the speed of sound can be written at first order in $\sigma_*$ as follows:
\be
s(\tau)=-\frac{1}{2}\sigma_*\beta_sF'\(-\beta_s\ln\tfrac{\tau}{\tau_f}\) +{\cal O}\(\sigma_*^2\)\ ,
\label{Fprimesharp}
\ee
where $'$ denotes the derivative with respect to the argument. Since we are considering sharp features happening around the time $\tau_f$, the functions involved in the integral \eref{GsrInt2} will only contribute for values in the neighborhood of $\tau_f$. Note that for polynomials with negative powers of $k\tau$, the approximation of evaluating them at $k\tau_f$ fails for small values of $k\tau$, since in that region they vary very rapidly. This may cause infrared divergences in the spectrum which, as we will see, can be cured by approximating the polynomials to first order around $k\tau_f$.

First, we define the variable $y\equiv -\beta_s\ln\left(\tau/\tau_f\right)$, and we expand the functions around $\tau=\tau_f$, which is equivalent to $y/\beta_s\ll 1$. Then, at first order, the expansion of $X$ in \eref{Xdef} reads:
\ba
X(kc_{s,0}\tau)\simeq\left[p_c\left(kc_{s,0}\tau_f\right)-y\frac{k\tau_f}{\beta_s}\frac{dp_c}{d(k\tau)}\bigg|_{\tau_f}\right]\cos\[2kc_{s,0}\tau_f\(1-\tfrac{y}{\beta_s}\)\]\hspace{2cm}\nn\\\nn\\
+\left[p_s\left(kc_{s,0}\tau_f\right)-y\frac{k\tau_f}{\beta_s}\frac{dp_s}{d(k\tau)}\bigg|_{\tau_f}\right]\sin\[2kc_{s,0}\tau_f\(1-\tfrac{y}{\beta_s}\)\]\ .
\label{Xexpanded}
\ea
Substituting in \eref{GsrInt2} the above expansion and the definition of $s$ \eref{Fprimesharp}, the change in the power spectrum is given by:  
\ba
\frac{\Delta{\cal P}_{{\cal R}}}{{\cal P}_{{\cal R},0}}=\frac{\sigma_*}{6}\left\{\Big[p_c\cos\left(2kc_{s,0}\tau_f\right)+p_s\sin\left(2kc_{s,0}\tau_f\right)\Big]\int_{-\infty}^{\infty} dy\,\cos\left(\frac{2kc_{s,0}\tau_f}{\beta_s}y\right)F'\left(y\right)\hspace{.45cm} \nn \right.\\\nn\\\left. +\Big[p_c\sin\left(2kc_{s,0}\tau_f\right)-p_s\cos\left(2kc_{s,0}\tau_f\right)\Big]\int_{-\infty}^{\infty} dy\,\sin\left(\frac{2kc_{s,0}\tau_f}{\beta_s}y\right)F'\left(y\right)\hspace{.45cm} \nn \right.\\\nn\\ \left.-\frac{k\tau_f}{\beta_s}\left[\frac{dp_s}{d(k\tau)}\bigg|_{\tau_f}\sin\left(2kc_{s,0}\tau_f\right)+\frac{dp_c}{d(k\tau)}\bigg|_{\tau_f}\cos\left(2kc_{s,0}\tau_f\right)\right]\int_{-\infty}^{\infty} dy\,\cos\left(\frac{2kc_{s,0}\tau_f}{\beta_s}y\right)y\,F'\left(y\right)\hspace{.45cm}\nn \right.\\\nn\\\left.+\frac{k\tau_f}{\beta_s}\left[\frac{dp_s}{d(k\tau)}\bigg|_{\tau_f}\cos\left(2kc_{s,0}\tau_f\right)-\frac{dp_c}{d(k\tau)}\bigg|_{\tau_f}\sin\left(2kc_{s,0}\tau_f\right)\right]\int_{-\infty}^{\infty} dy\,\sin\left(\frac{2kc_{s,0}\tau_f}{\beta_s}y\right)y\,F'\left(y\right)\right\}\ . \nn
\ea
Note that the integrals above are the Fourier transforms of the symmetric and antisymmetric parts of the derivative of the shape function $F$. We define the envelope functions resulting from these integrals as follows:
\ba
\int_{-\infty}^{\infty}dy\,\cos\left(\frac{2kc_{s,0}\tau_f}{\beta_s}y\right)F'(y)\equiv \frac{1}{2}\mathcal{D}_A\quad ,\qquad \int_{-\infty}^{\infty}dy\,\sin\left(\frac{2kc_{s,0}\tau_f}{\beta_s}y\right)F'(y)\equiv \frac{1}{2}\mathcal{D}_S\ ,\hspace{1.74cm}\label{FT1}\\\nn\\
\int_{-\infty}^{\infty}dy\,y\,F'(y)\cos\(\frac{2k c_{s,0}\tau_f}{\beta_s}y\)= \frac{\beta_s}{4c_{s,0}\tau_f}\frac{d}{dk}{\cal D}_S\quad ,\qquad\int_{-\infty}^{\infty}dy\,y\,F'(y)\sin\(\frac{2k c_{s,0}\tau_f}{\beta_s}y\)= -\frac{\beta_s}{4c_{s,0}\tau_f}\frac{d}{dk}{\cal D}_A\ ,\label{FT4}
\ea
where ${\cal D}_S$ and ${\cal D}_A$ are the envelope functions corresponding to the symmetric and antisymmetric parts of $F$, respectively. Finally, the change in the power spectrum can be written as:
\ba\label{eq:gsr3}
\frac{\Delta{\cal P}_{{\cal R}}}{{\cal P}_{{\cal R},0}}=\frac{\sigma_*}{12}\Bigg\{\Big[p_c\cos\left(2kc_{s,0}\tau_f\right)+p_s\sin\left(2kc_{s,0}\tau_f\right)\Big]\mathcal{D}_A+\Big[p_c\sin\left(2kc_{s,0}\tau_f\right)-p_s\cos\left(2kc_{s,0}\tau_f\right)\Big]\mathcal{D}_S \Bigg\} \nn \\\nn\\ 
-\frac{\sigma_*}{24c_{s,0}}\left\{\left[\frac{dp_s}{d(k\tau)}\bigg|_{\tau_f}\sin\left(2kc_{s,0}\tau_f\right)+\frac{dp_c}{d(k\tau)}\bigg|_{\tau_f}\cos\left(2kc_{s,0}\tau_f\right)\right]k\frac{d}{dk}\mathcal{D}_S\hspace{2cm} \right.\nn\\ \nn \\
\left.+\left[\frac{dp_s}{d(k\tau)}\bigg|_{\tau_f}\cos\left(2kc_{s,0}\tau_f\right)-\frac{dp_c}{d(k\tau)}\bigg|_{\tau_f}\sin\left(2kc_{s,0}\tau_f\right)\right]k\frac{d}{d k} \mathcal{D}_A \right\}
\ea
Let us stress that the contributions from the second and third lines are comparable to the ones in the first line. The infrared limit of the symmetric part is finite and tends to zero, which would not have been the case if we had only considered the zeroth order terms (first line). We will now substitute the values of the polynomials for the particular regime we are analyzing, $p_c=1/3$ and $p_s=-1/(3kc_{s,0}\tau)$. In this case, the change in the power spectrum reads:
\ba\label{eq:gsr4}
\frac{\Delta{\cal P}_{{\cal R}}}{{\cal P}_{{\cal R},0}}=\frac{\sigma_*}{36}\Bigg\{\Big[\cos\left(2kc_{s,0}\tau_f\right)-\frac{1}{kc_{s,0}\tau_f}\sin\left(2kc_{s,0}\tau_f\right)\Big]\mathcal{D}_A+\Big[\sin\left(2kc_{s,0}\tau_f\right)+\frac{1}{kc_{s,0}\tau_f}\cos\left(2kc_{s,0}\tau_f\right)\Big]\mathcal{D}_S \Bigg\} \nn \\\nn\\ 
-\frac{\sigma_*}{72}\Bigg\{\left[\frac{1}{(kc_{s,0}\tau_f)^2}\sin\left(2kc_{s,0}\tau_f\right)\right]k\frac{d}{dk}\mathcal{D}_S+\left[\frac{1}{(kc_{s,0}\tau_f)^2}\cos\left(2kc_{s,0}\tau_f\right)\right]k\frac{d}{d k} \mathcal{D}_A \Bigg\}\ .\hspace{1.3cm}
\ea


\subsubsection{Test for generic variations in the speed of sound}


\begin{figure}[t!]
\includegraphics[height=0.25\textheight,width=0.4\textwidth]{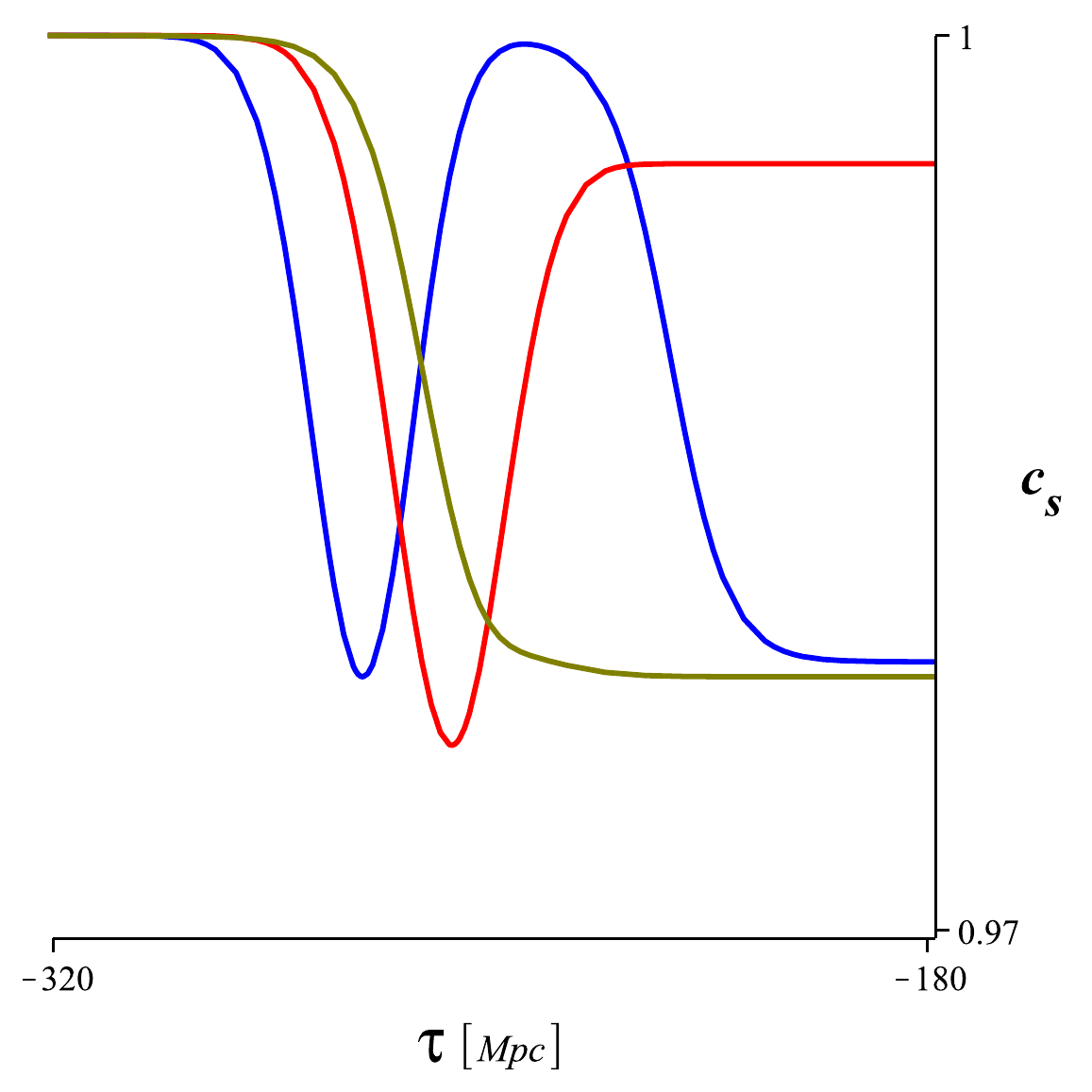}
\quad
\includegraphics[height=0.25\textheight,width=0.4\textwidth]{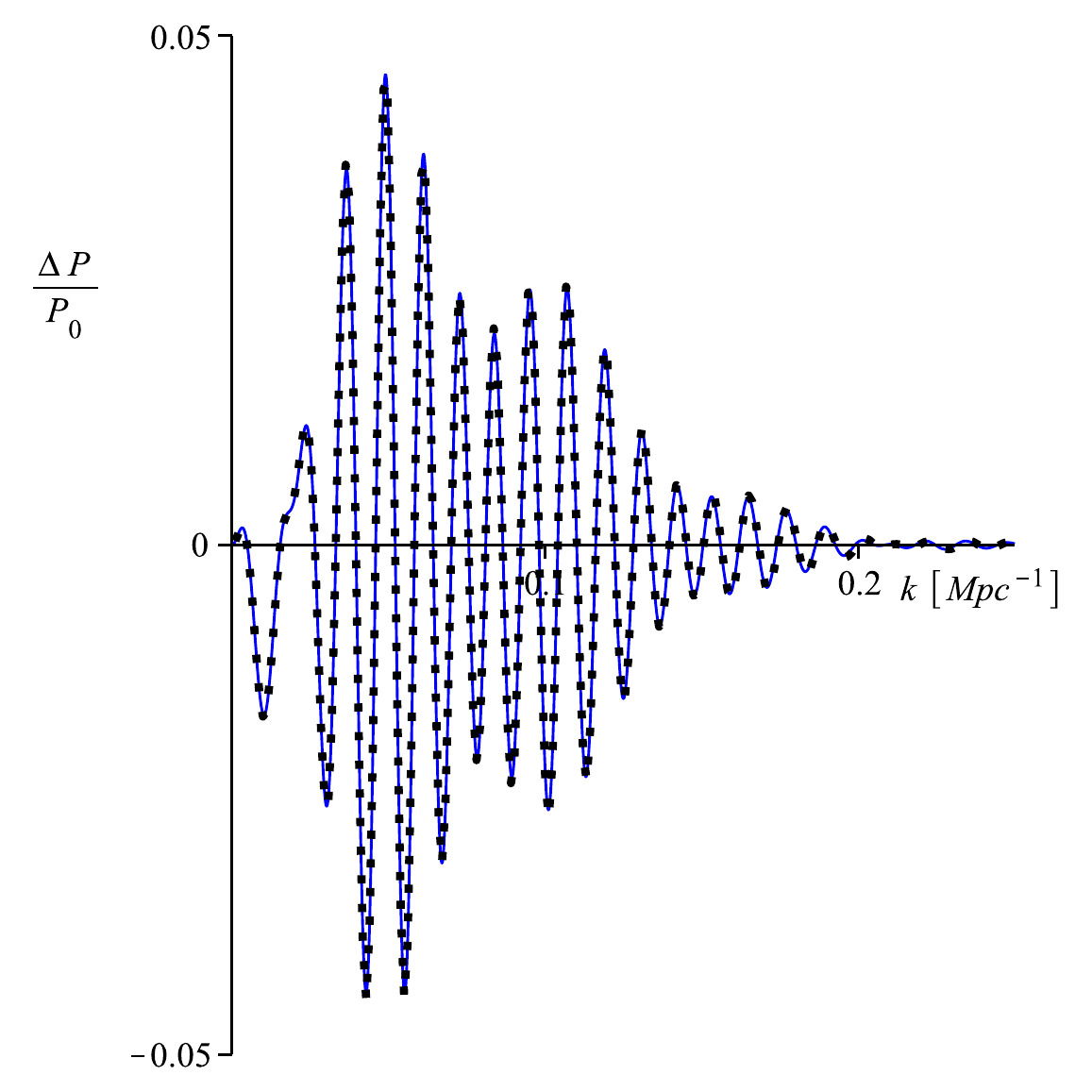}
\includegraphics[height=0.25\textheight,width=0.4\textwidth]{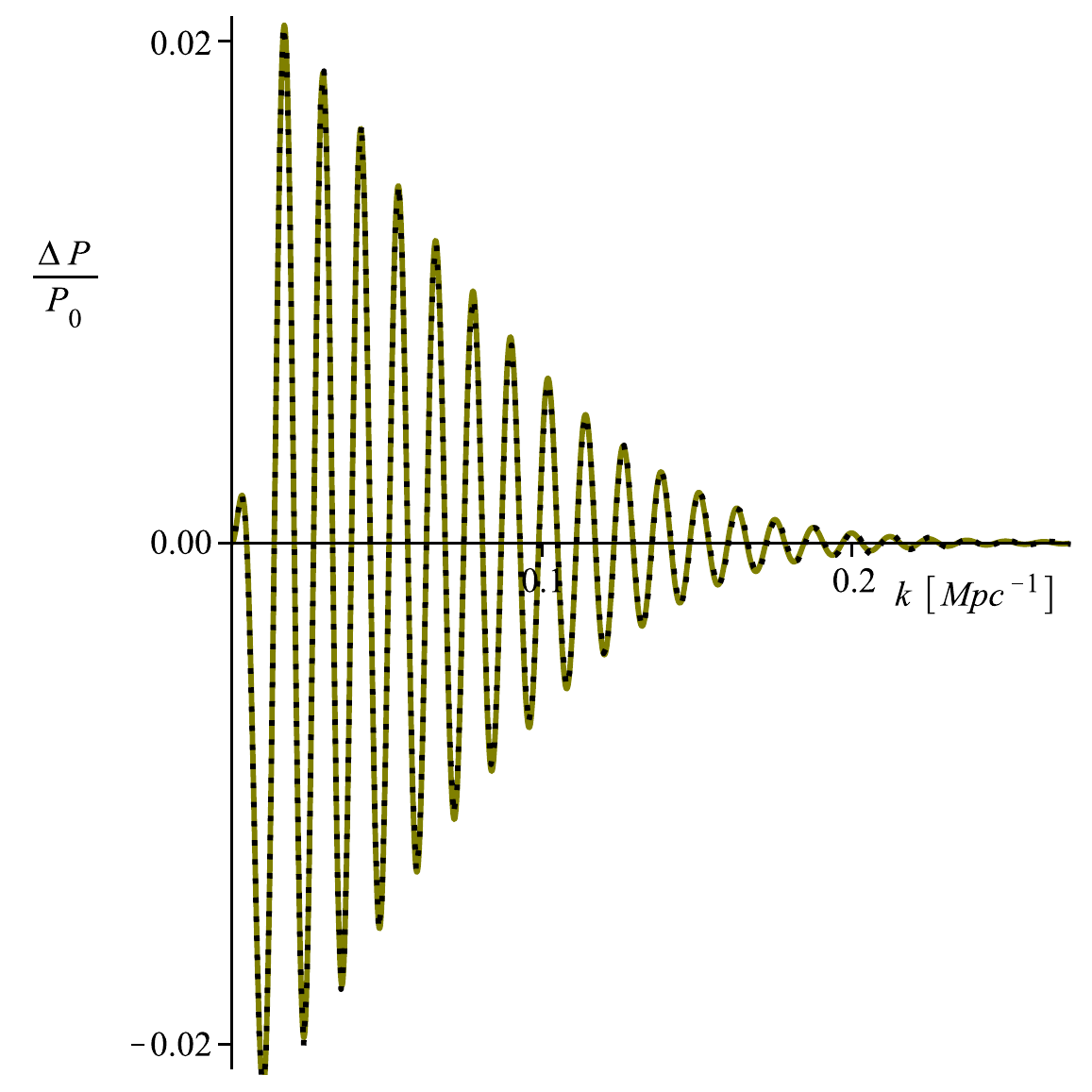}
\quad
\includegraphics[height=0.25\textheight,width=0.4\textwidth]{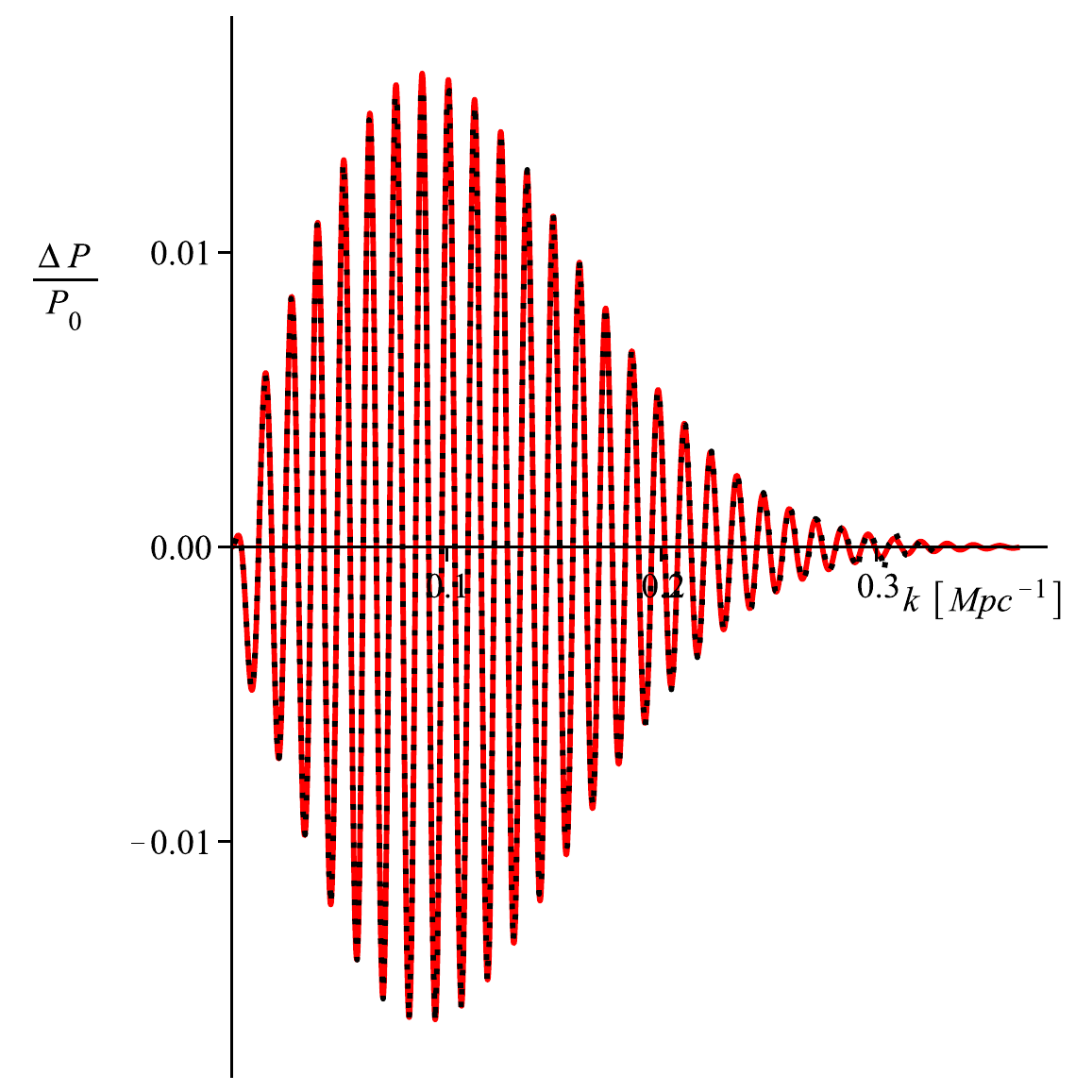}
\caption{Speed of sound as defined in (\ref{eq:cstot}) for three different values of the parameters. We show the power spectra calculated with the full integral (\ref{GsrInt2}) (dotted line) and with the approximation (\ref{eq:gsr3}) (solid line). The parameters, for the blue, olive and red figures, are respectively given by: $A=[-0.021,-0.0215,-0.0043]$, $B=[-0.043,-0.0086,-0.043]$, $\alpha^2=[\exp(6.3),\exp(6.3),\exp(7)]$, $\beta_s^2=[\exp(6.3),\exp(6.3),\exp(7)]$, $\tau_{0_g}=[-\exp(5.6),-\exp(5.55),-\exp(5.55)]$, $\tau_{0_t}=[-\exp(5.4),-\exp(5.55),-\exp(5.55)]$. For the first set of parameters the symmetric and antisymmetric parts have comparable magnitude, while for the second (third) set of parameters the antisymmetric (symmetric) part dominates. As can be seen by the very good agreement between the full integral and the approximation, the chosen parameters are all of them in the sharp feature regime.}
\label{fig:csd}
\end{figure}

\begin{figure}[ht!]
\includegraphics[height=0.16\textheight,width=0.23\textwidth]{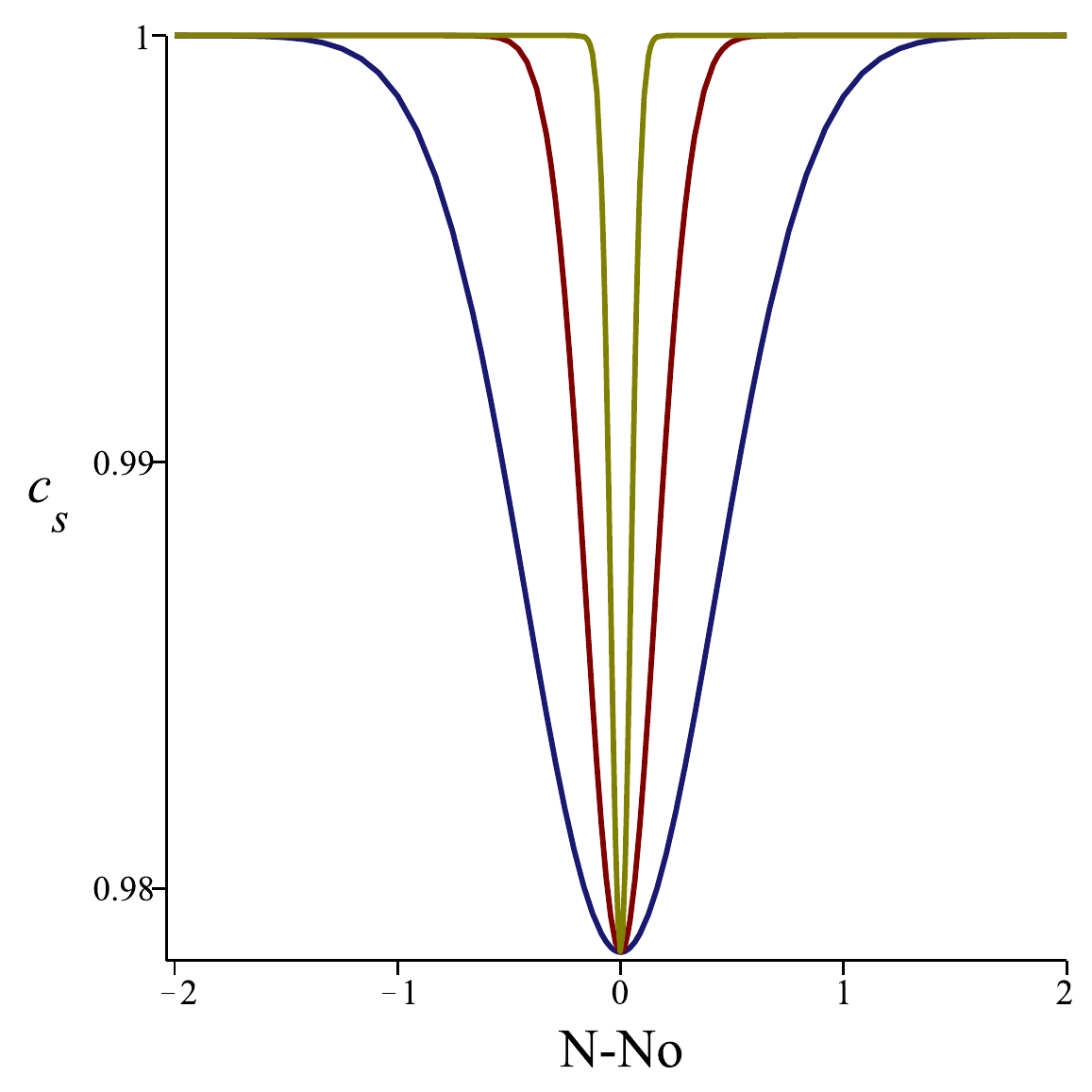}
\quad
\includegraphics[height=0.16\textheight,width=0.23\textwidth]{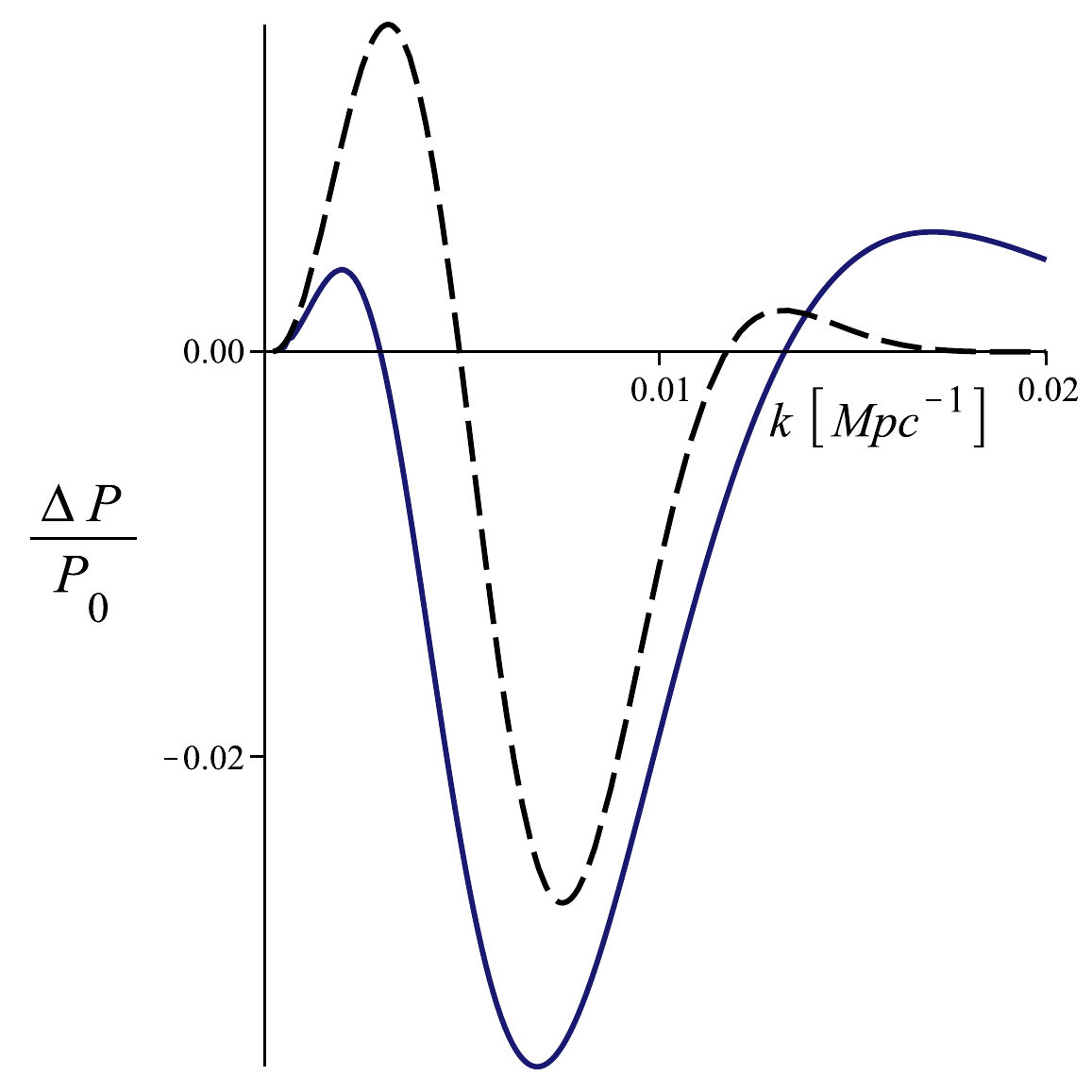}
\includegraphics[height=0.16\textheight,width=0.23\textwidth]{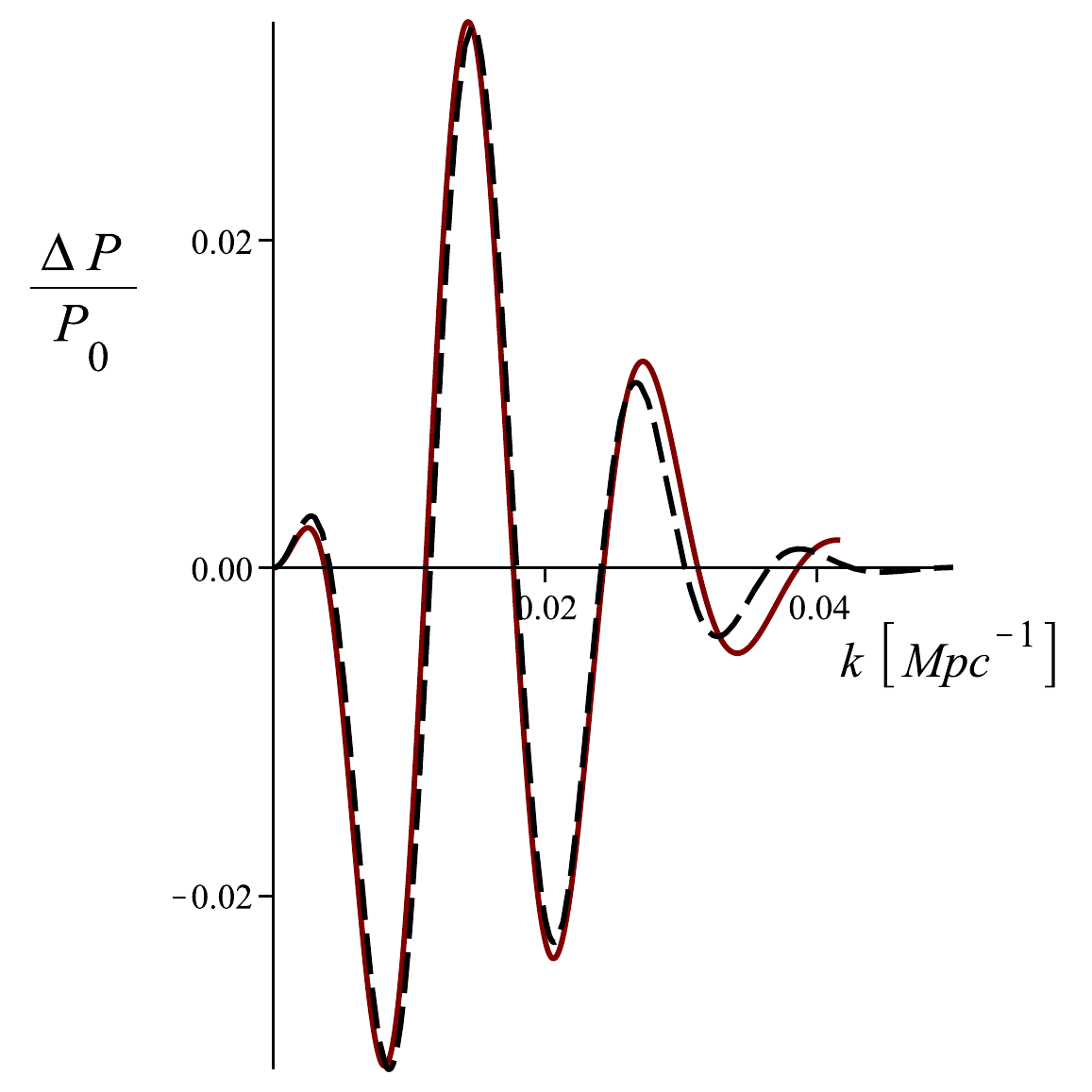}
\quad
\includegraphics[height=0.16\textheight,width=0.23\textwidth]{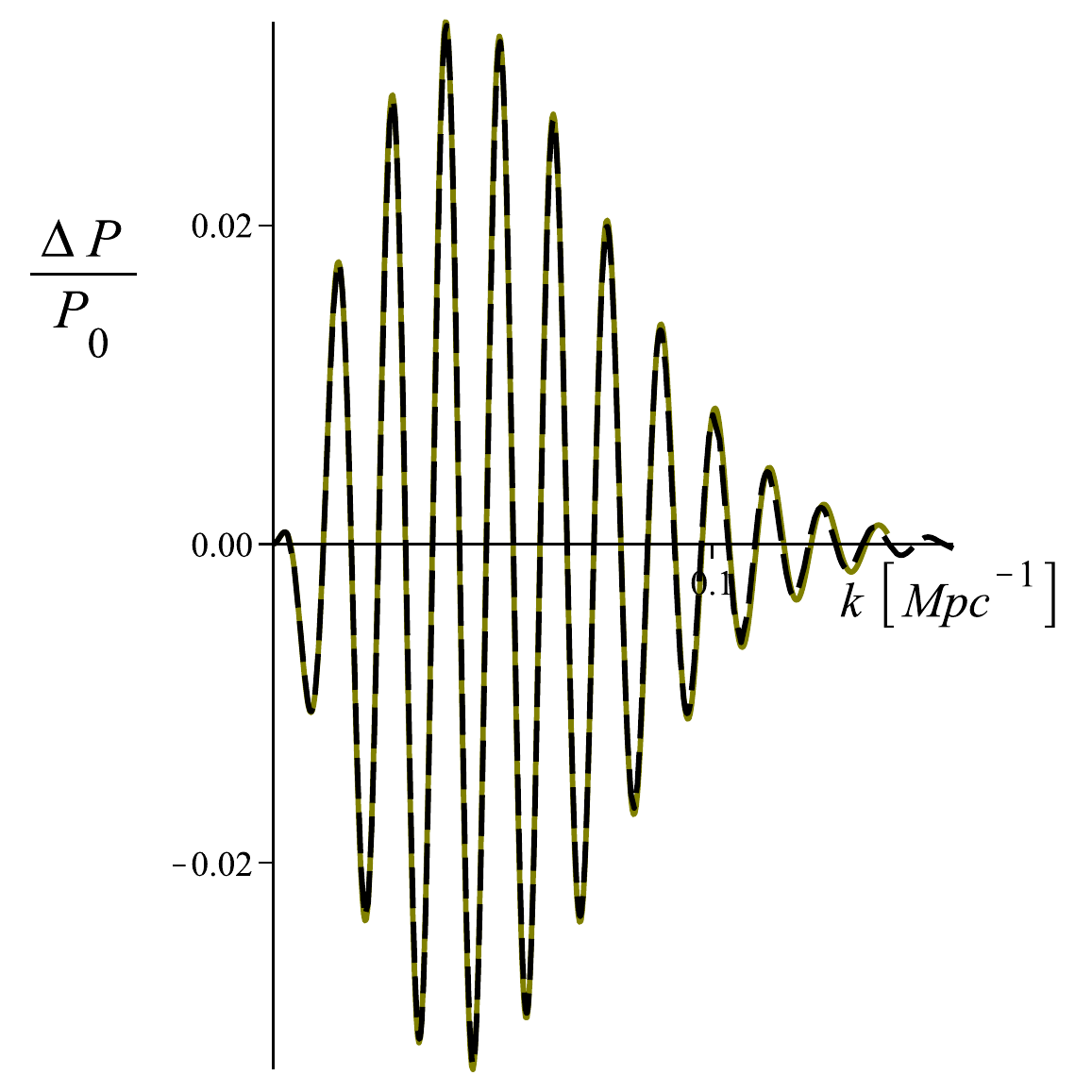}
\caption{Here we test when the approximation (\ref{eq:gsr3}) starts to break down. The full integral (\ref{GsrInt2}) is represented by dashed lines while the approximation  (\ref{eq:gsr3}) is given by solid lines. We take $A=0,\,B=-0.043,\,\tau_{0_g}=-\exp(5.55)$ for the three profiles of the speed of sound, and $\beta_g=[\exp(1),\,\exp(3),\,\exp(11/2)]$ for the blue, red and olive figures respectively. We see that the approximation starts to fail for features with $\Delta N\gtrsim 1$.} 
\label{fig:cds2}
\end{figure}

In this section we will test the approximation (\ref{eq:gsr3}) in comparison with the full integral (\ref{GsrInt2}). For the following particular example, we will explicitly decompose $c_s^2$ into its symmetric and antisymmetric parts:
\ba
c_s^2=1+A\left[1-\tanh\(\alpha\ln\tfrac{\tau}{\tau_{0_t}}\)\right]+B\exp\[{-\beta_s^2\(\ln\tfrac{\tau}{\tau_{0_g}}\)^2}\]\hspace{.8cm}\nn\\\nn\\
=\Bigg\{1+A+B\exp\[{-\beta_s^2\(\ln\tfrac{\tau}{\tau_{0_g}}\)^2}\]\Bigg\}_S+\Bigg\{-A\tanh\(\alpha\ln\tfrac{\tau}{\tau_{0_t}}\)\Bigg\}_A\ .\label{eq:cstot}
\ea
From the definitions \eref{Fsharp} and \eref{FT1} , the envelope functions are given by : 
\be
\mathcal{D}_A=-\frac{4\pi A}{\sigma_*}\frac{k\tau_{0_t}}{\alpha}\frac{1}{\sinh(\pi k\tau_{0_t}/\alpha)}\quad\quad,\quad\quad
\mathcal{D}_S=\frac{4\sqrt{\pi}B}{\sigma_*}\frac{k\tau_{0_g}}{\beta_s}\, \exp\(-\frac{k^2\tau_{0_g}^2}{\beta_s^2}\)\ .
\ee
Since the symmetric and antisymmetric parts do not necessarily peak at the same time, the integrands involved in each part take values around $\tau_{0_g}$ and $\tau_{0_t}$, respectively. We test our approximation for different values of the parameters above, and show our results in figure \ref{fig:csd}. We can see that the approximation is indeed very good, and that it allows to reproduce highly non-trivial power spectra. By allowing $\beta_s$ and/or $\alpha$ to be small, we can see where the approximation starts to fail. We show these results in figure \ref{fig:cds2}, where one can see that for features with $\Delta N\gtrsim 1$ the approximation breaks down.


\subsection{C. Comparison of power spectra}


\begin{figure}[b!]
\begin{minipage}[h]{3in}
\includegraphics[width=2.7in]{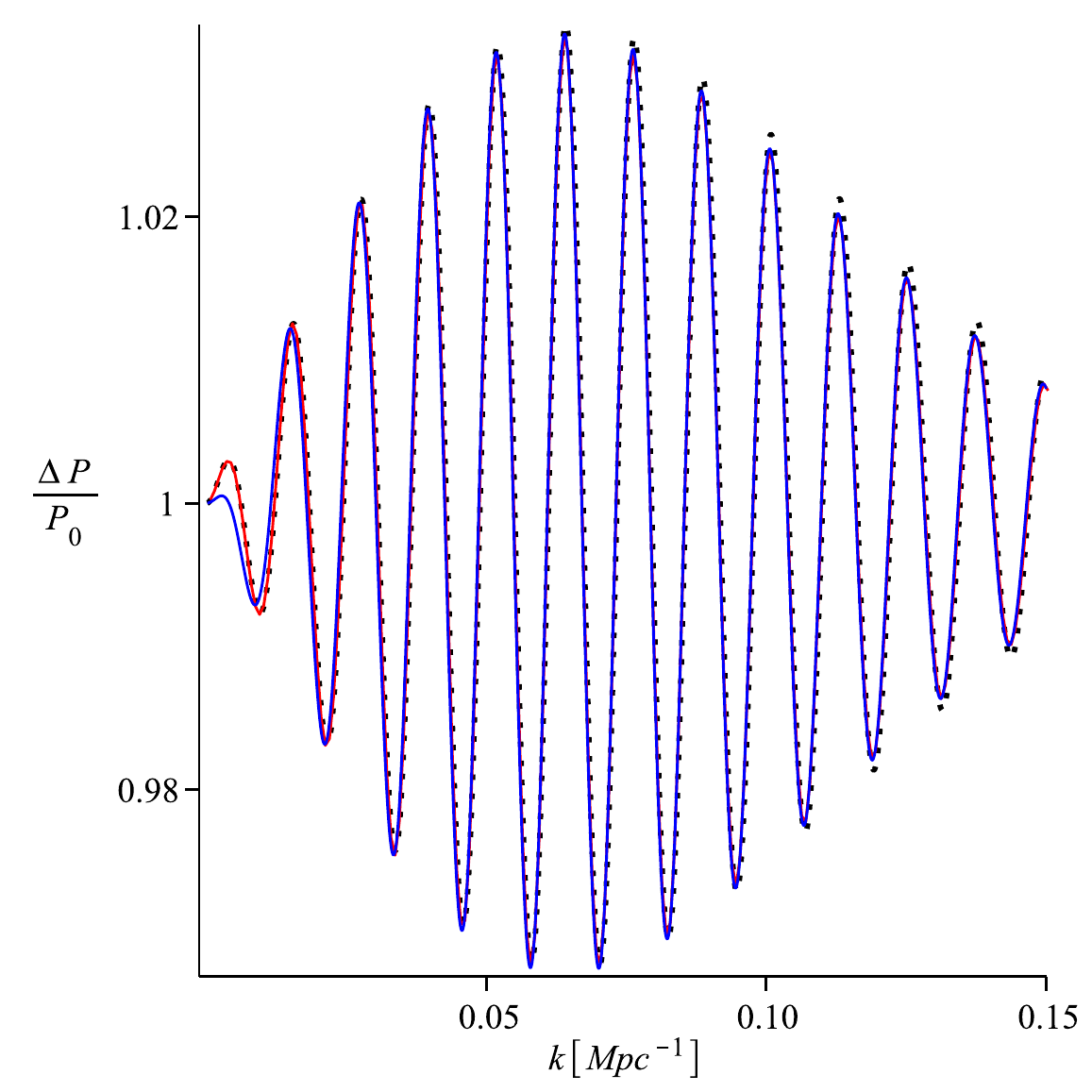}
\end{minipage}
\begin{minipage}[h]{3in}
\includegraphics[width=2in]{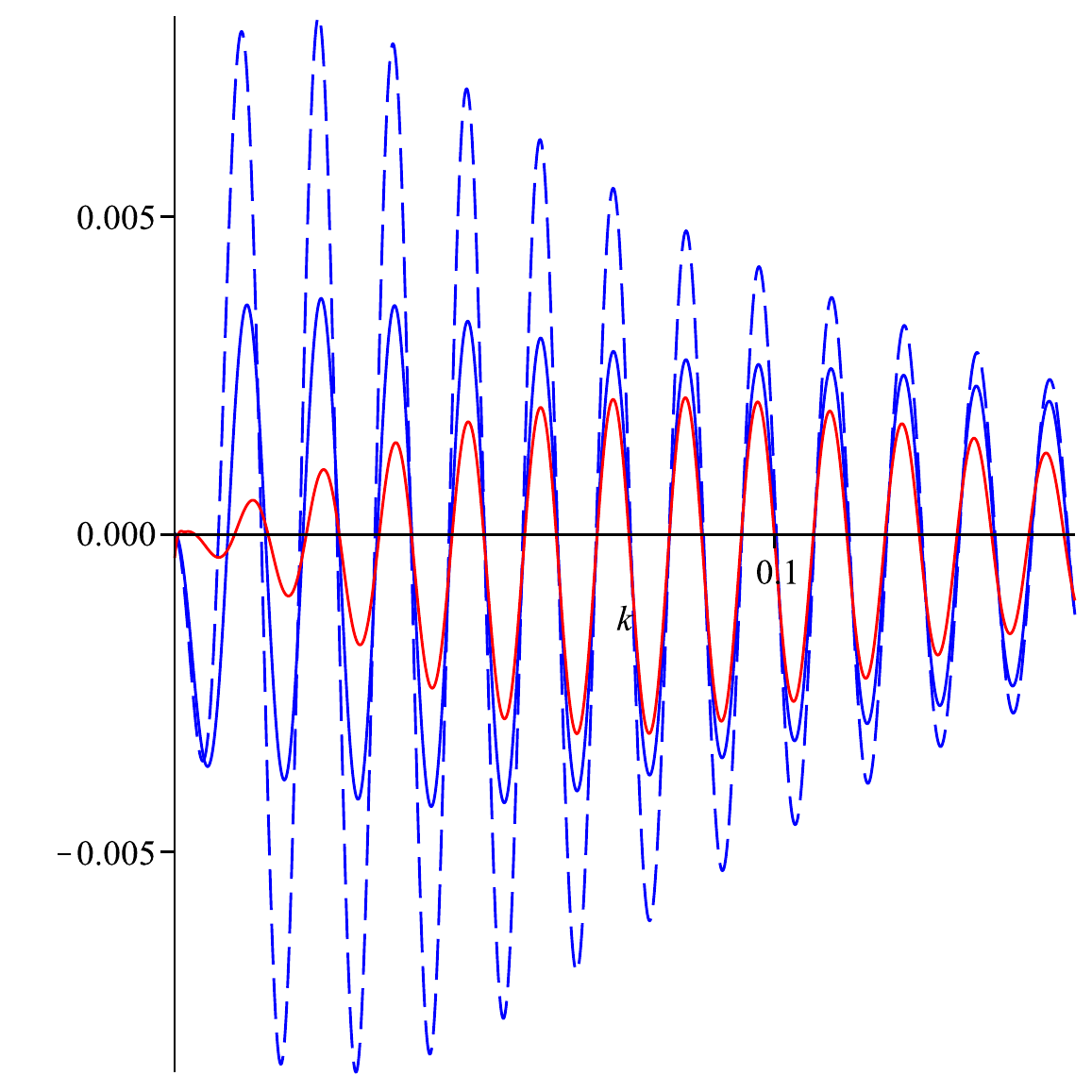}
\end{minipage}
\caption{Change in the power spectrum due to a reduced speed of sound given by \eref{eq:gaussefolds2}, with the following choice of parameters: $B=-0.043$, $\beta_s=23.34$, $\ln(\tau_f)=5.55$, corresponding to one of our best fits to the Planck CMB power spectrum \cite{Achucarro:2013cva}. LEFT: different methods to compute the primordial power spectrum: GSR in the sharp feature approach (blue), SRFT (red), and a solution obtained from the numerical solution to the mode equation \eref{MSequation} (black dotted). RIGHT: differences of the GSR sharp feature method (solid blue) and SRFT (red) against the numerical solution. The dashed blue line is the GSR sharp feature approach if we had not taken into account the term proportional to $s$ in the source function \eref{sourcesharp}. The numerical solution is calculated choosing $\epsilon\simeq1.25\times 10^{-4}$ and $\tilde\eta\simeq -0.02$. Higher values of $\epsilon$ need a proper accounting for the slow-roll corrections.}
\label{fig:methods}
\end{figure}


In this section we apply both SRFT and GSR methods for moderately sharp reductions to calculate the change in the power spectrum, and compare them with the power spectrum calculated from the numerical solution to the mode equation \eref{MSequation}. We will test a reduction in the speed of sound purely symmetric in the variable $y=-\beta_s\ln (\tau/\tau_f)$:
\begin{equation}
u=1-c_s^{-2}=B\,e^{-\beta_s^2(N-N_f)^2}=B\,e^{-\beta_s^2\left(\ln\frac{\tau}{\tau_f}\right)^2}\ .
\label{eq:gaussefolds2}
\end{equation}
In figure \ref{fig:methods} we show the comparison between the power spectrum coming from the GSR result \eref{eq:gsr3} with the one coming form the SRFT method (\ref{eq:deltappfourier}), and with a numerical solution. In general terms, both methods are in good agreement with the numerical solution. We also note that at large scales the SRFT method reproduces the numerical results better than the GSR method. This is partly due to the fact that in the GSR approximation we have only taken a subset of the terms in the source function. The agreement would have been much worse if we had not taken into account the term proportional to $s$, as the dashed line in the right plot of figure \ref{fig:methods} indicates.
Note that $k\tau_f\sim 1$ corresponds to the first peak in the left plot of fig. \ref{fig:methods} above, precisely the regime where we expect a discrepancy, as anticipated in eq. \eref{GSRtoSRFT}.

This shows that, in the regime of moderately sharp variations of the speed of sound, the simple SRFT formula \eref{eq:deltappfourier} is capable of reproducing the effect of \emph{all} the terms in the equation of motion, and that there is no need to impose any further hierarchy between the different terms of the equation of motion in order to have a simple expression, as long as slow-roll is uninterrupted.


\subsection{D. Bispectrum for moderately sharp reductions}


In this section we will compute the change in the bispectrum due to moderately sharp reductions in the speed of sound using the in-in formalism. Instead of the SRFT method reviewed in \S II A, we will use an approximation based on sharp features \cite{Bartolo:2013exa}, as for the power spectrum. Our starting point is the cubic action in the effective field theory of inflation, where we will only take into account the contribution from variations in the speed of sound at first order:
\be
S_3=\int d^4x\, a^3M_\text{Pl}^2\frac{\epsilon}{H}\left\{2Hsc_s^{-2}{\cal R}\dot{\cal R}^2+\(1-c_s^{-2}\)\dot{\cal R}\[\dot{\cal R}^2-\frac{1}{a^2}\(\nabla{\cal R}\)^2\]\right\}\ ,\label{s3sharp}
\ee
with ${\cal R}=-\pi H$. For sharp features $(\beta_s\gg1)$ and given the parametrization in \eref{Fsharp} and \eref{Fprimesharp}, one is tempted to think that the contribution of $s$ will dominate over the contribution of $(1-c_s^{-2})$. However, we will show that the contributions arising from both terms are of the same order, {\it independently of the sharpness} $\beta_s$. As dictated by the in-in formalism, the three-point correlation function reads:
\ba
\la{\cal R}_{{\bf k}_1}{\cal R}_{{\bf k}_2}{\cal R}_{{\bf k}_3}\ra=\bigg\la\Re\bigg\{2i\,{\cal R}_{{\bf k}_1}(0){\cal R}_{{\bf k}_2}(0){\cal R}_{{\bf k}_3}(0)\int_{-\infty}^0d\tau\int d^3x\, a^4M_\text{Pl}^2\frac{\epsilon}{H}\[2Hsc_s^{-2}{\cal R}\dot{\cal R}^2\right.\hspace{3cm}\label{bi1}\\\nn\\\left.+\(1-c_s^{-2}\)\dot{\cal R}^3-H^2\tau^2\(1-c_s^{-2}\)\dot{\cal R}\(\nabla{\cal R}\)^2\]\bigg\}\bigg\ra\ ,\hspace{1cm}\nn
\ea
where we have used that\footnote{Note that the expression $a=-1/(H\tau)$ is only valid for uninterrupted slow-roll. In the case of slow-roll violations, especially for sharp steps in the potential, the corrections may give additional contributions to the correlation functions.} $a=-1/(H\tau)$. After expressing the functions ${\cal R}(\tau,{\bf x})$ in Fourier space and using the Wick theorem, we obtain
\ba
\la{\cal R}_{{\bf k}_1}{\cal R}_{{\bf k}_2}{\cal R}_{{\bf k}_3}\ra=\Re\bigg\{2i\, u_{{\bf k}_1}(0)u_{{\bf k}_2}(0)u_{{\bf k}_3}(0)\int_{-\infty}^0\frac{d\tau}{\tau^2}\frac{\epsilon M_\text{Pl}^2}{H^2}(2\pi)^3\int d^3q_1\int d^3q_2\int d^3q_3\,\delta({\bf q}_1+{\bf q}_2+{\bf q}_3)\times\hspace{1.8cm}\label{bi2}\\\nn\\\times\[4sc_s^{-2}u_{{\bf q}_1}^*(\tau)u_{{\bf q}_2}^{*\prime}(\tau)u_{{\bf q}_3}^{*\prime}(\tau)\Big(\delta({\bf k}_1-{\bf q}_1)\delta({\bf k}_2-{\bf q}_2)\delta({\bf k}_3-{\bf q}_3)+\{{\bf k}_1\leftrightarrow {\bf k}_2\}+\{{\bf k}_1\leftrightarrow {\bf k}_3\}\Big)\right.\hspace{3.5cm}\nn\\\nn\\\left. -6\tau\(1-c_s^{-2}\)u_{{\bf q}_1}^{*\prime}(\tau)u_{{\bf q}_2}^{*\prime}(\tau)u_{{\bf q}_3}^{*\prime}(\tau)\delta({\bf k}_1-{\bf q}_1)\delta({\bf k}_2-{\bf q}_2)\delta({\bf k}_3-{\bf q}_3)\right.\hspace{6.7cm}\nn\\\nn\\\left. -2\tau\(1-c_s^{-2}\)\({\bf q}_2\cdot {\bf q}_3\)u_{{\bf q}_1}^{*\prime}(\tau)u_{{\bf q}_2}^{*}(\tau)u_{{\bf q}_3}^{*}(\tau)\Big(\delta({\bf k}_1-{\bf q}_1)\delta({\bf k}_2-{\bf q}_2)\delta({\bf k}_3-{\bf q}_3)+\{{\bf k}_1\leftrightarrow {\bf k}_2\}+\{{\bf k}_1\leftrightarrow {\bf k}_3\}\Big)\]\bigg\}\hspace{0.2cm}\ .\nn
\ea
For the leading order contribution, it suffices to use the zeroth-order mode function
\be
u_{\bf k}(\tau)=\frac{iH}{\sqrt{4\epsilon c_{s,0}k^3}}\(1+ikc_{s,0}\tau\)e^{-ikc_{s,0}\tau}\ ,
\label{modefunctions}
\ee
and the three-point correlation function is then:
\ba
\la{\cal R}_{{\bf k}_1}{\cal R}_{{\bf k}_2}{\cal R}_{{\bf k}_3}\ra=\frac{{\cal P}_{{\cal R},0}^2(2\pi)^7M_\text{Pl}^6}{8k_1^3k_2^3k_3^3}\delta({\bf k}_1+{\bf k}_2+{\bf k}_3)\int_{-\infty}^0d\tau\,\bigg\{\cos\(Kc_{s,0}\tau\)\Big[4sc_s^{-2}c_{s,0}^3\tau k_1k_2k_3(k_1k_2+\text{\footnotesize [2 perm]})\hspace{2cm}\label{bi3}\\\nn\\-2\tau c_{s,0}\(1-c_s^{-2}\)\big[k_1^2(k_2+k_3)({\bf k}_2\cdot{\bf k}_3)+\text{\footnotesize [2 perm]}\big]\Big]-\sin\(Kc_{s,0}\tau\)\Big[4sc_s^{-2}c_{s,0}^2 (k_1^2k_2^2+\text{\footnotesize [2 perm]})\hspace{2cm}\nn\\\nn\\-6\tau^2 c_{s,0}^4\(1-c_s^{-2}\)k_1^2k_2^2k_3^2-2\(1-c_s^{-2}\)\big[k_1^2({\bf k}_2\cdot{\bf k}_3)+\text{\footnotesize [2 perm]}\big]+2\tau^2c_{s,0}^2\(1-c_s^{-2}\)k_1k_2k_3\big[k_1({\bf k}_2\cdot{\bf k}_3)+\text{\footnotesize [2 perm]}\big]\Big]\bigg\}\ ,\nn
\ea
where $K\equiv k_1+k_2+k_3$ and\footnote{Notice that the definition of ${\cal P}_{{\cal R},0}$ in \S IIA did not include $c_{s,0}$, since in the SRFT approach it is taken to be one.} ${\cal P}_{{\cal R},0}=H^2/(8\pi^2\epsilon M_\text{Pl}^2c_{s,0})$. Before we proceed, some comments are in order:

\begin{itemize}

\item For steps in the potential, one also has to calculate the contribution to the three-point function coming from similar cubic operators. It is easy to track the polynomials in $k_i$ arising from the different operators if one pays attention to the form of the mode functions \eref{modefunctions}. This way, we noticed that the result for steps in the potential in \cite[eq.\ 3.32]{Bartolo:2013exa} is missing a term, so it should display as follows:
\ba
\frac{{\cal G}}{k_1k_2k_3}=\frac{1}{4}\epsilon_\text{step}{\cal D}\(\frac{K\tau_f}{2\beta}\)\[\(\frac{k_1^2+k_2^2+k_3^2}{k_1k_2k_3\tau_f}-K\tau_f\)K\tau_f\cos(K\tau_f)\right.\\\nn\\\left. -\(\frac{k_1^2+k_2^2+k_3^2}{k_1k_2k_3\tau_f}-\frac{\sum_{i\neq j}k_i^2k_j}{k_1k_2k_3}K\tau+K\tau\)\sin(K\tau_f)\]\nn
\ea
This is indeed good news, since the missing term $(+K\tau)$ above was the source of a small discrepancy found by the authors of \cite{Bartolo:2013exa} with respect to previous results \cite{Adshead:2011jq}, of order $10-15\%$ on large scales. We have checked that this discrepancy vanishes when the extra term is introduced.

\item We consider sharp features ($\beta_s\gg1$) peaking in $\tau_f$ and define the new variable $y$ through $\tau=\tau_f\,e^{-y/\beta_s}$, as we did for the power spectrum. There are two kinds of functions appearing in \eref{bi3}: polynomials and oscillating functions. For the latter, we substitute $\tau\simeq\tau_f(1-y/\beta_s)$ and do not expand further, in order to keep the Fourier transforms. For the former, the zeroth order approximation $\tau\simeq\tau_f$ (as in \cite{Bartolo:2013exa}) provides excellent results,\footnote{As opposed to the power spectrum, in this case we only have polynomials with positive powers of $k\tau$, and therefore evaluating them at $k\tau_f$ is already a good approximation for sufficiently sharp features.} although we take the next order and evaluate them at $\tau\simeq\tau_f(1-y/\beta_s)$ to test for not-so-sharp features. We will therefore calculate the first order correction to previous results. Furthermore we consider, apart from the operator ${\cal R}\dot{\cal R}^2$ (proportional to $s$), two extra contributions $\dot{\cal R}^3$ and $\dot{\cal R}(\nabla{\cal R})^2$ (proportional to $u$) and show that they all contribute at the same order, independently of the sharpness $\beta_s$. This is because, although $s$ is proportional to the sharpness $\beta_s$, it is also proportional to the derivative of the shape function, $F'$, defined in eq.\ \eqref{Fprimesharp}. On the other hand, $u$ is proportional to the shape function, but the Fourier transform of $F$ introduces an additional factor $\beta_s$ relative to the Fourier transform of $F'$, cf. eqs. \eref{FT1},\eref{FT4} and \eref{transform1}--\eref{transform4}.

\item The integrals in \eref{bi3} contain Fourier transforms of the shape function $F$ and its derivative, given the definitions in \eref{Fsharp} and \eref{Fprimesharp}. The symmetric and antisymmetric envelope functions arising from the Fourier transform of $F'$ were already defined in \eref{FT1} and \eref{FT4}. For completeness, we will give the complementary definitions obtained when integrating by parts:
\ba
\int_{-\infty}^{\infty}dy\,F(y)\cos\(\frac{K c_{s,0}\tau_f}{\beta_s}y\)= -\frac{\beta_s}{2Kc_{s,0}\tau_f}{\cal D}_S\quad ,\qquad\int_{-\infty}^{\infty}dy\,F(y)\sin\(\frac{K c_{s,0}\tau_f}{\beta_s}y\)= \frac{\beta_s}{2Kc_{s,0}\tau_f}{\cal D}_A\ ,\quad\label{transform1}\\\nn\\
\int_{-\infty}^{\infty}dy\,y\,F(y)\cos\(\frac{K c_{s,0}\tau_f}{\beta_s}y\)= \frac{1}{2}\(\frac{\beta_s}{Kc_{s,0}\tau_f}\)^2\(K\frac{d{\cal D}_A}{dK}-{\cal D}_A\)\ ,\label{transform3}\quad\hspace{2.5cm}\\\nn\\
\int_{-\infty}^{\infty}dy\,y\,F(y)\sin\(\frac{K c_{s,0}\tau_f}{\beta_s}y\)= \frac{1}{2}\(\frac{\beta_s}{Kc_{s,0}\tau_f}\)^2\(K\frac{d{\cal D}_S}{dK}-{\cal D}_S\)\ ,\quad\hspace{2.55cm}\label{transform4}
\ea
where the slight change of notation between these definitions and those in \eref{FT1} and \eref{FT4} is given by $K\leftrightarrow 2k$. We also imposed that $F$ asymptotically vanishes when integrating by parts, which will be the case in this calculation.

\end{itemize}

Taking into account the comments above, we calculate the bispectrum to leading order \eref{bi3} for the particular case in which $c_{s,0}=1$, so that we can compare to the SRFT method described in \S IIA. We will express the bispectrum in terms of the normalized scale-dependent function $f_\text{NL}({\bf k}_1,{\bf k}_2,{\bf k}_3)$ defined by:
\be
\la{\cal R}_{{\bf k}_1}{\cal R}_{{\bf k}_2}{\cal R}_{{\bf k}_3}\ra=(2\pi)^3\delta({\bf k}_1+{\bf k}_2+{\bf k}_3)\Delta B_{\cal R}=(2\pi)^7\delta({\bf k}_1+{\bf k}_2+{\bf k}_3)\frac{3}{10}f_\text{NL}({\bf k}_1,{\bf k}_2,{\bf k}_3){\cal P}_{{\cal R},0}^2M_{\text{Pl}}^6\frac{k_1^3+k_2^3+k_3^3}{k_1^3k_2^3k_3^3}\ ,
\label{binormalization}
\ee
and we will use the following identities for a triangle of vectors $\{{\bf k}_1,{\bf k}_2,{\bf k}_3\}$: 
\ba
k_1({\bf k}_2\cdot {\bf k}_3)+\text{\footnotesize [2 perm]}&=&\frac{1}{2}\[k_1^3+k_2^3+k_3^3-K(k_1k_2+\text{\footnotesize [2 perm]})+3k_1k_2k_3\]\ ,\\
k_1^2({\bf k}_2\cdot {\bf k}_3)+\text{\footnotesize [2 perm]}&=&\frac{1}{2}\[k_1^4+k_2^4+k_3^4-2(k_1^2k_2^2+\text{\footnotesize [2 perm]})\]\ ,\nn\\
k_1^2(k_2+k_3)({\bf k}_2\cdot {\bf k}_3)+\text{\footnotesize [2 perm]}&=&\frac{1}{2}\[K(k_1^4+k_2^4+k_3^4)-(k_1^5+k_2^5+k_3^5)-K(k_1^2k_2^2+\text{\footnotesize [2 perm]})-k_1k_2k_3(k_1k_2+\text{\footnotesize [2 perm]})\]\ .\nn
\ea
Finally, the bispectrum contribution due to variations in the speed of sound as considered in the cubic action \eref{s3sharp}, to first order in the size of the feature $\sigma_*$, and to first order in the polynomial expansion $\tau\simeq\tau_f(1-y/\beta_s)$ reads:
\ba
f_\text{NL}({\bf k}_1,{\bf k}_2,{\bf k}_3)=\frac{5}{24}\frac{\sigma_*}{k_1^3+k_2^3+k_3^3}\Bigg\{\cos\(K\tau_f\)\bigg\{\tau_f^2\,\frac{k_1k_2k_3}{K}\Big[(k_1^3+k_2^3+k_3^3)+K(k_1k_2+\text{\footnotesize [2 perm]})-3k_1k_2k_3\Big]{\cal D}_A\hspace{1.2cm}\label{completebi}\\\nn\\
+\frac{\tau_f}{K}\Big[K(k_1^4+k_2^4+k_3^4)-(k_1^5+k_2^5+k_3^5)+K(k_1^2k_2^2+\text{\footnotesize [2 perm]})-4k_1k_2k_3(k_1k_2+\text{\footnotesize [2 perm]})\hspace{1.2cm}\nn\\\nn\\
+3\frac{k_1k_2k_3}{K}(k_1^3+k_2^3+k_3^3)-9\frac{k_1^2k_2^2k_3^2}{K}\Big]{\cal D}_S-3\tau_f\frac{k_1k_2k_3}{K}\Big[(k_1^3+k_2^3+k_3^3)+\frac{1}{3}K(k_1k_2+\text{\footnotesize [2 perm]})-3k_1k_2k_3\Big]\frac{d{\cal D}_S}{dK}\hspace{1.2cm}\nn\\\nn\\
-\frac{1}{K^2}\Big[3K(k_1^4+k_2^4+k_3^4)-2(k_1^5+k_2^5+k_3^5)-4K(k_1^2k_2^2+\text{\footnotesize [2 perm]})-2k_1k_2k_3(k_1k_2+\text{\footnotesize [2 perm]})\Big]{\cal D}_A\hspace{1.2cm}\nn\\\nn\\
+\frac{1}{K}\Big[2K(k_1^4+k_2^4+k_3^4)-2(k_1^5+k_2^5+k_3^5)-2k_1k_2k_3(k_1k_2+\text{\footnotesize [2 perm]})\Big]\frac{d{\cal D}_A}{dK}\hspace{1.2cm}\nn\\\nn\\
-\frac{1}{\tau_fK^2}\Big[(k_1^4+k_2^4+k_3^4)-2(k_1^2k_2^2+\text{\footnotesize [2 perm]})\Big]\({\cal D}_S-K\frac{d{\cal D}_S}{dK}\)\bigg\}
+\sin\(K\tau_f\)\bigg\{\{{\cal D}_S\leftrightarrow{\cal D}_A\ ,\ \tau_f\leftrightarrow-\tau_f\}
\bigg\}\Bigg\}\ ,\hspace{1.2cm}\nn
\ea
where the $\sin(K\tau_f)$ in the last line contains the same terms as the $\cos(K\tau_f)$, but changing ${\cal D}_S\leftrightarrow{\cal D}_A$ and $\tau_f\leftrightarrow-\tau_f$, as indicated. This is the formula we want to compare with \eref{bispectrumana}, after proper normalization. Below, we show the comparison for different functional forms of the speed of sound.



\subsection{E. Comparison of bispectra}

\begin{figure}[t!]
\includegraphics[scale=.31]{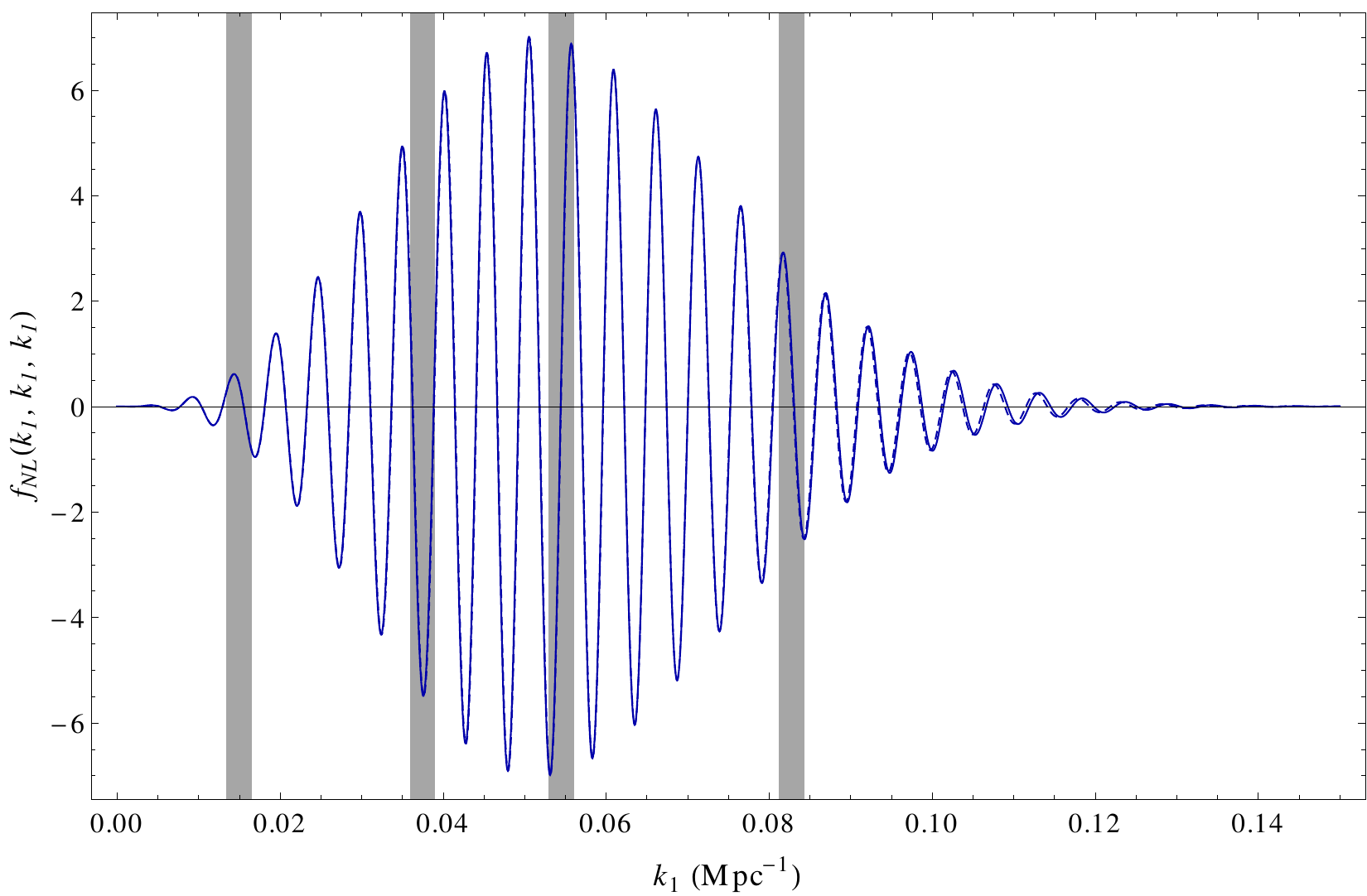}
\quad
\includegraphics[scale=.31]{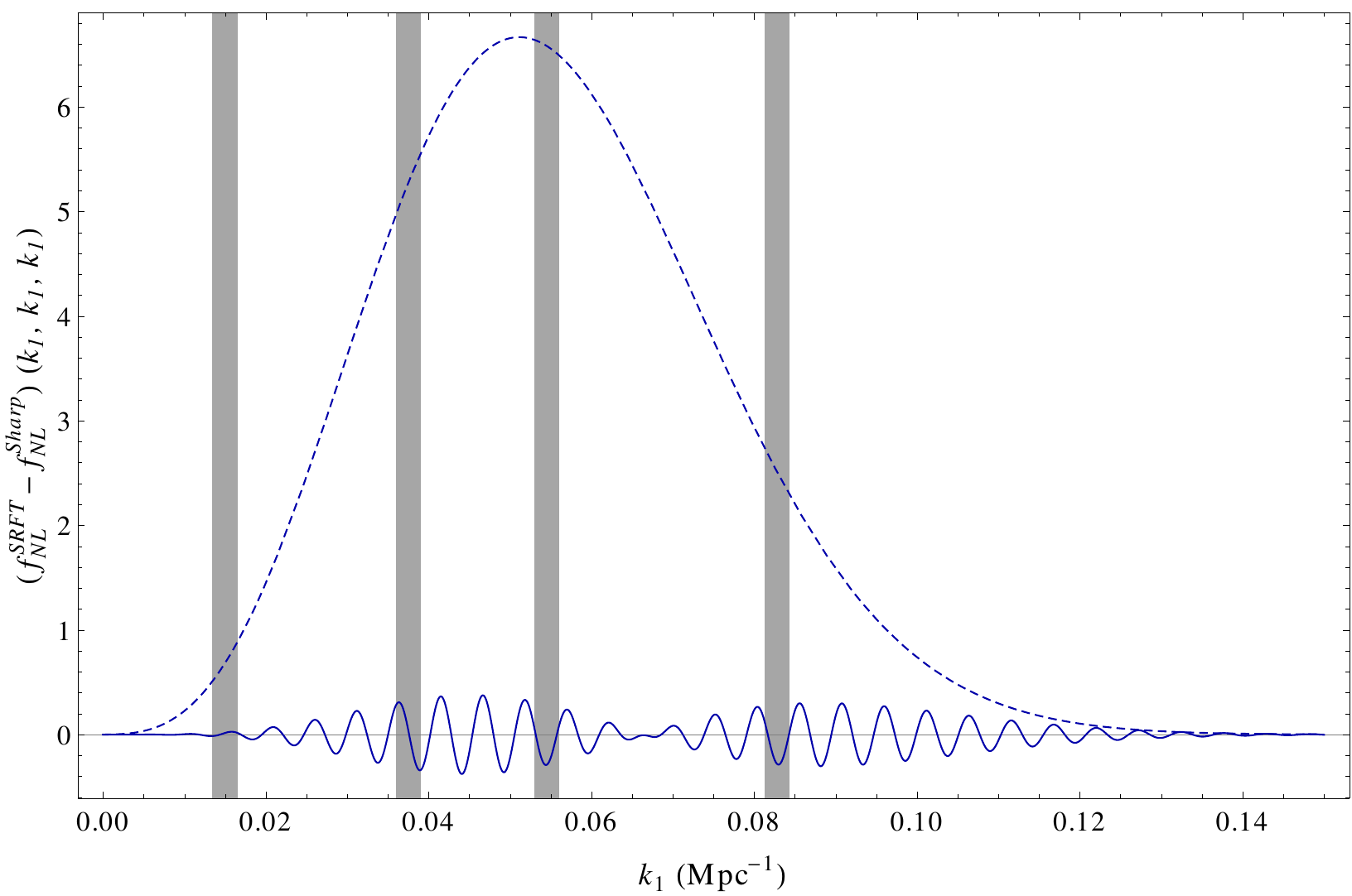}
\includegraphics[scale=.31]{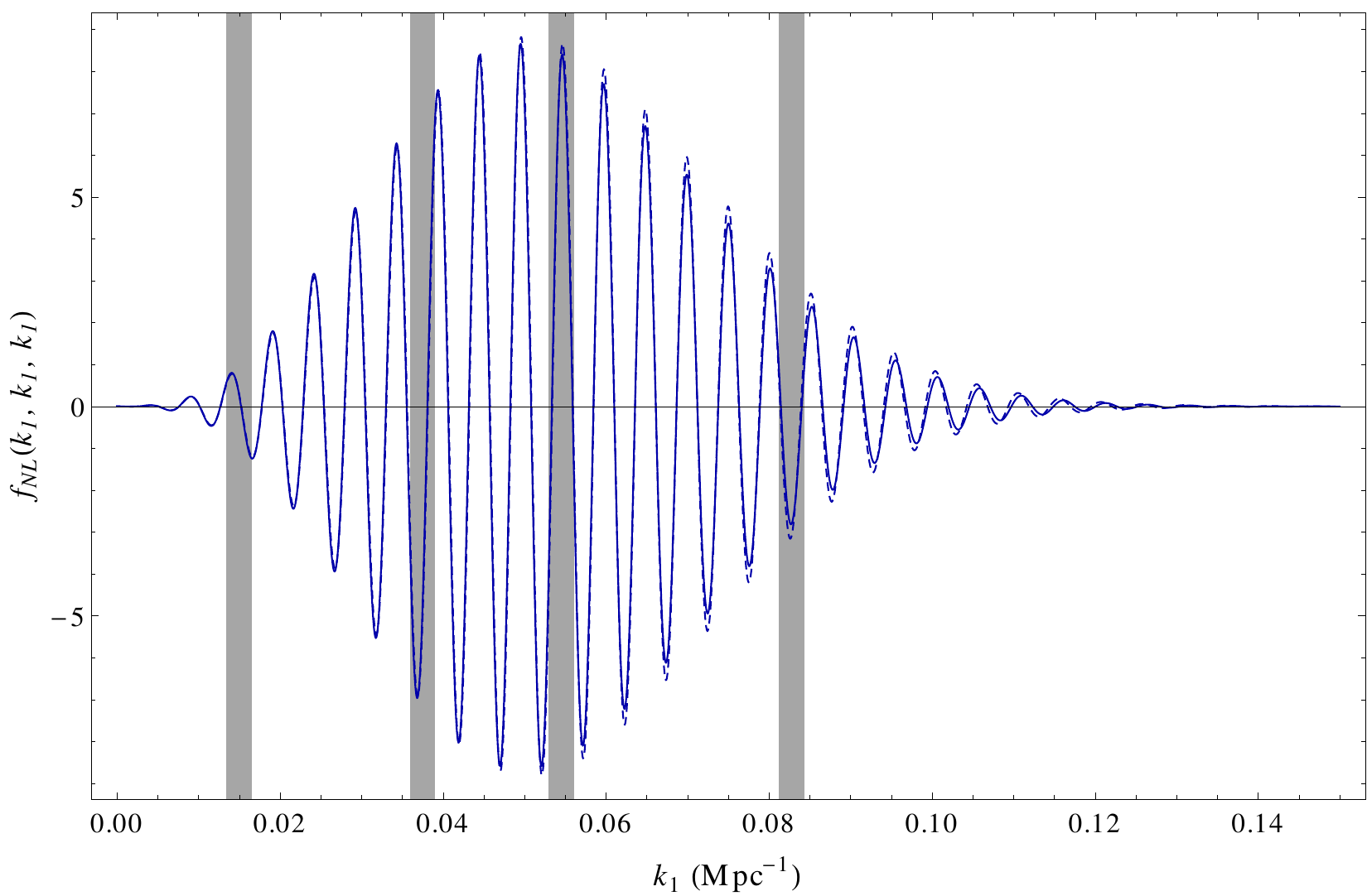}
\quad
\includegraphics[scale=.31]{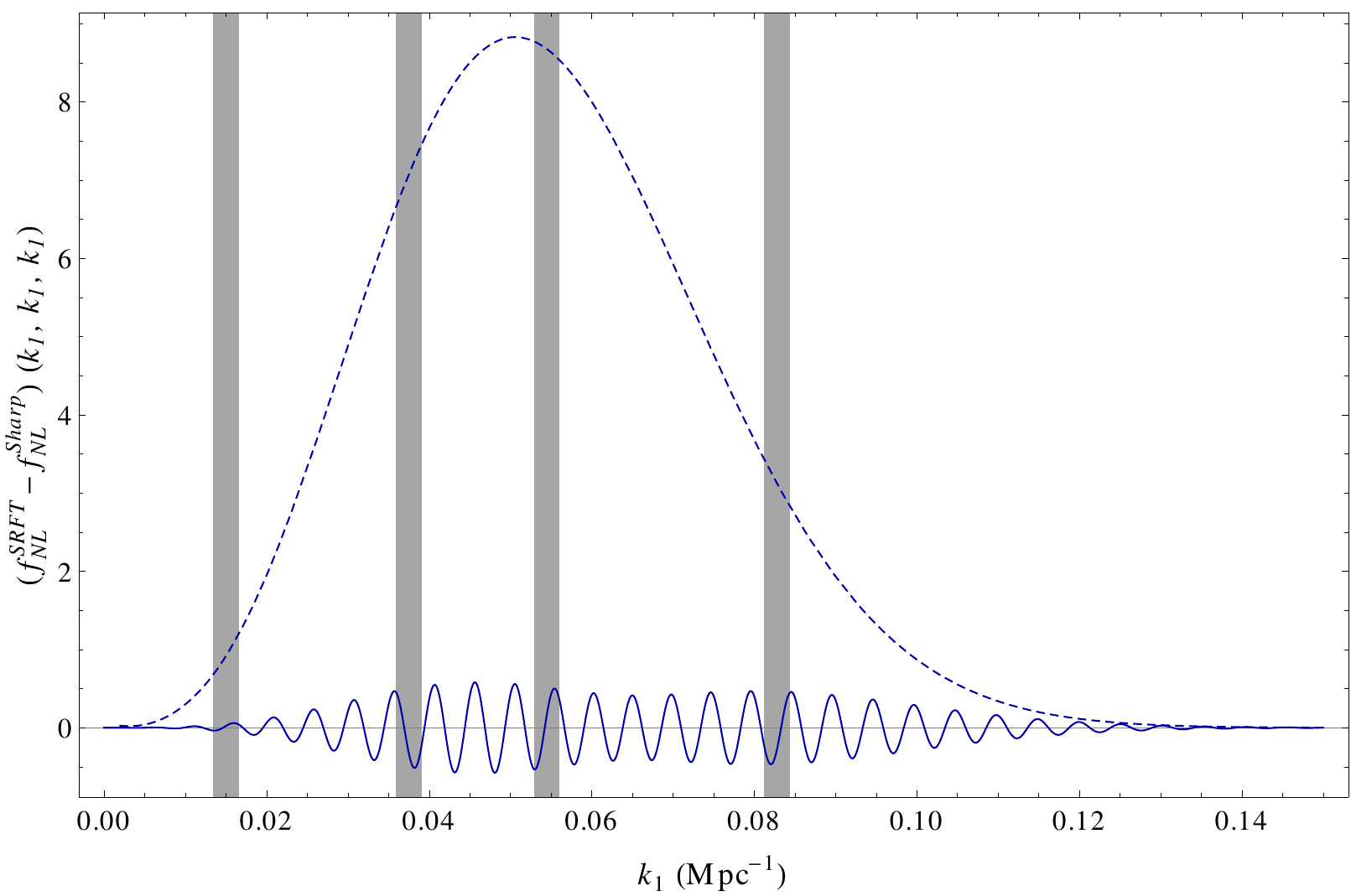}

\caption{LEFT: bispectrum $f_\text{NL}$ signal in the equilateral limit with the normalization indicated in \eref{binormalization}, given by a symmetric reduction in the speed of sound as in \eref{sharpgauss} (TOP) and an asymmetric reduction as in \eref{sharpanti} (BOTTOM), calculated with the SRFT formula \eref{bispectrumana} (solid) and with the sharp approximation \eref{completebi} (dashed). RIGHT: absolute difference between the signals showed in the left plot (solid), together with the envelope of the signal (dashed). The grey strips represent the approximate scales of the first four acoustic peaks of the CMB temperature spectrum. The parameters are $\sigma_*=0.04$, $\beta_s=25.5$, $\ln (-\tau_f)=6$. This gives $|s|_\text{max}\simeq 0.42$ for the symmetric case and $|s|_\text{max}\simeq 0.55$ for the asymmetric case. Note that in both cases the relative difference with respect to the envelope is large only at very small scales, which in any case will be indistinguishable at the observational level. We are also within the limit $|s|_\text{max}<1$, where these signatures are reliable but sharp enough so that the sharp approximation works.}
\label{bispectrumcomparison}
\end{figure}

In this section we compare the bispectrum obtained using the SRFT method \eref{bispectrumana} with that using the first order approximation for sharp features \eref{completebi}. As a first example, one can reproduce our test case of gaussian reductions in the speed of sound, cf. \eref{eq:gaussefolds}, by taking:
\be
F=\exp\[-\beta_s^2\(\ln\tfrac{\tau}{\tau_f}\)^2\]\quad\Rightarrow\quad 1-c_s^{-2}=-\sigma_*\, e^{-\beta_s^2\(\ln\tfrac{\tau}{\tau_f}\)^2}+{\cal O}\(\sigma_*\)^2\ ,
\label{sharpgauss}
\ee
where the correspondence between this set of parameters and the one used in \cite{Achucarro:2013cva} is $\sigma_*\leftrightarrow -B$, $\tau_f\leftrightarrow\tau_0$, and $\beta_s\leftrightarrow\sqrt{\beta}$. In this case $F$ is symmetric in the variable $y=-\beta_s\ln\tfrac{\tau}{\tau_f}$ and therefore only the symmetric envelope function ${\cal D}_S$ contributes, which is given by
\be
{\cal D}_S=-\frac{2K\tau_f}{\beta_s}\sqrt{\pi}\exp\(-\frac{K^2\tau_f^2}{4\beta_s^2}\)\quad ,\quad {\cal D}_A=0\ .
\ee
In figure \ref{bispectrumcomparison} we show the excellent agreement between the results obtained with \eref{bispectrumana} and \eref{completebi} for the equilateral limit $k_1=k_2=k_3$. We have checked that for other configurations in momentum space, such as the folded or the squeezed shapes, the agreement is very similar. Note that in figure \ref{bispectrumcomparison} we are plotting the absolute difference in $f_\text{NL}$ and comparing with the total envelope of the signal.\footnote{We point out that the total envelope of the signal is not given by ${\cal D}_S$ or ${\cal D}_A$ alone. The total envelope is a combination of both functions, their derivatives, and the polynomials of $k_i$ that appear in \eref{completebi}.} At small scales one can see that the relative difference compared to the total signal is high, due to the fact that the approximation for sharp features starts to fail for large values of $K\tau$. However, the absolute signal is insignificant at such small scales. 

As a second example, we propose a shape function with an antisymmetric part:
\be
F=\exp\[-\beta_s^2\(\ln\tfrac{\tau}{\tau_f}\)^2+\beta_s\ln\tfrac{\tau}{\tau_f}\]\quad\Rightarrow\quad 1-c_s^{-2}=-\sigma_*\(\frac{\tau}{\tau_f}\)^{\beta_s}\, e^{-\beta_s^2\(\ln\tfrac{\tau}{\tau_f}\)^2}+{\cal O}\(\sigma_*\)^2\ ,
\label{sharpanti}
\ee
for which the symmetric and antisymmetric envelope functions read:
\be
{\cal D}_S=-\frac{2K\tau_f}{\beta_s}\sqrt{\pi}\exp\(\frac{\beta_s^2-K^2\tau_f^2}{4\beta_s^2}\)\cos\(\frac{K\tau_f}{2\beta_s}\)\quad ,\quad {\cal D}_A=-\frac{2K\tau_f}{\beta_s}\sqrt{\pi}\exp\(\frac{\beta_s^2-K^2\tau_f^2}{4\beta_s^2}\)\sin\(\frac{K\tau_f}{2\beta_s}\)
\ee
We show in figure \ref{bispectrumcomparison} the equilateral bispectrum signal produced by the asymmetric shape \eref{sharpanti}, again derived using \eref{bispectrumana} and \eref{completebi}. As one can see in figure \ref{bispectrumcomparison}, the agreement is also remarkable for functions with an antisymmetric part.


\section{III.  Parameter space and details of the search} 

In our previous work \cite{Achucarro:2013cva} we proposed a test case consisting of a gaussian reduction in the speed of sound. The functional form is inspired by soft turns along a multi-field inflationary trajectory with a large hierarchy of masses, a situation that is consistently described by an effective single-field theory and uninterrupted slow roll \cite{Achucarro:2010jv,Achucarro:2010da,Cespedes:2012hu,Achucarro:2012sm,Gao:2012uq,Gao:2013ota}. We parametrized the reduction in the speed of sound as a gaussian in e-folds $N$ of inflation:
\begin{equation}
u=1-c_s^{-2}=B\,e^{-\beta(N-N_0)^2}=B\,e^{-\beta\left(\ln\frac{\tau}{\tau_0}\right)^2}\ ,
\label{eq:gaussefolds}
\end{equation}
where $\beta>0$ is the sharpness, $B<0$ is the amplitude, and $N_0$ (or $\tau_0$) is the instant of maximal reduction. Assuming slow-roll, the conformal time $\tau$ is related to the e-folds of inflation through $\ln\left(-\tau\right)=\left(N_\text{in}-N\right)-\ln\left(a_\text{in}H_0\right)$, where $a_\text{in}=a(N_\text{in})$ and $N_\text{in}$ is the time when the last $\sim$ 60 e-folds of inflation start. Notice that the quantity $N_\text{in}$ is irrelevant, since all the quantities in e-folds are defined with respect to $N_\text{in}$.


\subsection{A. Choice of parameter space}


There are two main criteria that we followed in order to determine the parameter regions that we explored:
\begin{enumerate}[\bf (a)]
\item
The angular scales probed by Planck ($\ell=2-2500$) roughly correspond to certain momentum scales crossing the Hubble sound horizon during the first $N_\text{CMB}\simeq$ 7 e-folds of the last $\sim$ 60 e-folds of inflation. If the data resembles features due to a reduced speed of sound, it is most likely to find them in this ``CMB window'', so we choose to `look under the lamppost'. This means that the sharpness $\beta$ and the position $N_0$ are chosen so that the reduction happens well within this window. As a by-product, we avoid degeneracies with the spectral index $n_s$ and the optical depth $\tau_\text{reio}$ due to very wide reductions.\footnote{
Note that the \emph{lamppost} is actually bigger, since any feature happening in a particular window propagates in the primordial power spectra to a bigger region. E.g.\ modes that leave the horizon after the reduction in $c_s$ has finished are also affected by it. Thus, it would be interesting to extend our search to larger values of $|\tau_0|$.}
\item
The SRFT calculation of the power spectrum and the bispectrum is valid for mild and moderately sharp reductions of the speed of sound. Also, the slow-roll contributions to the bispectrum are disregarded with respect to the terms arising from the reduced speed of sound \cite{Achucarro:2012fd}. This means that the amplitude $|u|$ and the rate of change $s\equiv \tfrac{\dot c_s}{c_sH}$ must be much smaller than one, while being (at least one of them) much larger than the slow-roll parameters. As a bonus, later in the text we will argue that $|s|\ll1$ is tightly related to an adiabatic evolution \cite{Cespedes:2012hu}.
\end{enumerate}

We took a very conservative definition for the total width of the reduction (in e-folds): ten standard deviations, $\Delta N={10}/{\sqrt{2\beta}}$. Then, from {\bf (a)}, the position $N_0$ and the sharpness $\beta$ should satisfy ${5}{\sqrt{2\beta}}<N_0< N_\text{CMB}-{5}{\sqrt{2\beta}}$ and ${10}{\sqrt{2\beta}}<N_\text{CMB}$. As to the perturbative regime, the rate of change $s$ of the speed of sound \eref{eq:gaussefolds} reads:
\begin{equation}\label{eq:s_gaussN}
s(N)=\frac{dc_s}{c_sdN}=-\frac{B\beta(N-N_0)\,e^{-\beta(N-N_0)^2}}{1-B\,e^{-\beta(N-N_0)^2}}\ .
\end{equation}
Since we have to impose $|s|\ll1$ for all values of $N$, it suffices to impose this condition at the point where $|s|$ takes its maximum value $|s(N_*)|=|s|_\text{max}$, determined by:
\begin{equation}
N_*=N_0\pm\frac{1}{\sqrt{2\beta}}\sqrt{1+{\cal O}(B)}\ \simeq \ N_0\pm\frac{1}{\sqrt{2\beta}}\ ,
\end{equation} 
which approximately corresponds to one standard deviation of our gaussian, and we have used that $|B|\ll1$. Then the condition $|s|_\text{max}\ll1$ translates into $\beta\ll\tfrac{2e}{B^2}+{\cal O}(B^{-1})$. Altogether, the allowed region of our parameter space is taken to be \cite{Achucarro:2013cva}:
\begin{subequations}
\label{eq:bounds}
\begin{gather}
{\cal O}(\epsilon,\eta)    \ll|B|\ll         1\ ,\label{eq:bound1}\\
\frac{50}{N_\text{CMB}^2}<\beta\ll         \frac{2e}{B^2}\ ,\label{eq:bound2}\\
\frac{5}{\sqrt{2\beta}}    <N_0< N_\text{CMB}-\frac{5}{\sqrt{2\beta}}\ .\label{eq:bound3}
\end{gather}
\end{subequations}
Notice that, as explained above in {\bf (b)}, the bound $|B|\gg{\cal O}(\epsilon,\eta)$ can be avoided if $|s|_\text{max}\gg{\cal O}(\epsilon,\eta)$. For computational purposes, we use the parameter $\ln(-\tau_0)$ instead of $N_0$ for the data analysis. The range for this parameter is taken to be more strongly restricted than by \eref{eq:bound3}:
\begin{equation}
4.4 \leq \ln(-\tau_0) \leq 6\ ,\label{eq:2bound3}
\end{equation}
The features in the power spectrum and bispectrum are linearly oscillating, as well as those tested in one of the searches for bispectrum features by the Planck collaboration \cite[sec.\ 7.3.3]{Ade:2013ydc}. The oscillatory frequency is determined by $\tau_0$, and the range of frequencies covered in Planck's bispectrum analysis is equivalent to the interval $\ln(-\tau_0)\in\[4.43,5.34\]$, which motivated us to search in the interval given above. Hence, our search is slightly larger than theirs in this respect.


\subsection{B. Perturbative unitarity and adiabatic evolution}


In the recent works \cite{Adshead:2014sga,Cannone:2014qna}, consistency conditions regarding inflationary models that produce features were studied. In particular they derive several bounds from the requirement that the theory describing the features is in the weak coupling regime. In this section we clarify what these bounds mean in the context of soft transient reductions in the speed of sound, in particular for our test case \cite{Achucarro:2013cva}.

In \cite{Cannone:2014qna}, they establish a hard upper bound on the sharpness of the feature, based on the loss of unitarity when the loop contribution to a correlation function becomes of the same order as the tree level correlator:\footnote{This calculation is possible thanks to the fact that for the case of a feature in the Hubble parameter the $n$-order lagrangian acquires a particularly simple form \cite{Behbahani:2011it}.} $\beta_\text{CBM}\lesssim 160$, where $\beta_\text{CBM}$ (labelled by the initials of the authors of \cite{Cannone:2014qna}) defines the sharpness of the feature: $\beta_\text{CBM}\equiv 1/(H\Delta t)$.

Our sharpness parameter $\beta$ is related to that of \cite{Cannone:2014qna} by $\beta=50\beta_\text{CBM}^2$, where we took the conservative definition of the width to be ten standard deviations, as explained in \S IIIA. This imposes the following bound on our sharpness parameter:
\begin{equation}
\ln\beta\lesssim 14\ .
\label{boundAH}
\end{equation}
Since we restricted our search to $2<\ln\beta <7.5$, the fits we found in that region \cite{Achucarro:2013cva} are perfectly consistent with the bound given above. Even if we take the crude definition for the width of only one standard deviation, the correspondence would be $\beta=\beta_\text{CBM}^2$, and the bound would translate to $\ln\beta\lesssim 10$, which still leaves us in a safe region. The analysis of \cite{Adshead:2014sga} goes along the same lines as that of \cite{Cannone:2014qna}, and similar results are obtained. They also identify additional scales above which the theory breaks down.
Given that we \emph{a priori} constrained our search to a region of the parameter space where the perturbative and adiabatic regimes are respected, it remains by far within the bounds derived in \cite{Adshead:2014sga,Cannone:2014qna}, and therefore the predictions obtained are consistently interpreted by the underlying theory.

It was also found \cite{Adshead:2014sga,Cannone:2014qna} that the best fit so far for steps in the potential in the CMB \cite{Ade:2013uln,Benetti:2013cja,Miranda:2013wxa} does not lie within the allowed theoretical bounds. This calls into question the consistency of the framework in which these predictions are derived. More interestingly, this motivates a new theoretical framework able to consistently describe those predictions, since the data is blind to whether a theory is internally consistent or not.

An important  and evident conclusion of these analyses is that very sharp features are problematic from the theoretical point of view. In addition, one could speculate that if the data finally points to inflationary scenarios with large field excursions, a (slightly broken) symmetry should protect the background, and then we would not expect to find sharp features in the potential. This further motivates the study of moderately sharp features, which are still safely described by an underlying theory.

The previous results were obtained in the framework of the effective field theory of inflation \cite{Cheung:2007st} taking into account only the time dependence of the Hubble parameter, and neglecting the variation of the rest of coefficients $M_n^4$. First of all, it is not clear whether similar conclusions would hold considering changes in the $M_n^4$ coefficients. It is possible to construct the $n$-order lagrangian for the case of changes in the Hubble parameter and group all the terms together in a single vertex (for example $\pi^n$) by successive integration by parts. However, this is in general very difficult for changes in the $M_n^4$ coefficients, since the number of degrees of freedom is larger. In the absence of a UV theory that gives us a recipe for consistently calculating $M_n^4$, any estimate on how they determine the perturbative regime must be made with extreme caution. 
  
Last but not least, the intuition in terms of scattering amplitudes is borrowed from the standard QFT techniques which assume time-independent vertex coefficients. Intuitively, this will be applicable to time-dependent coefficients if they obey an adiabatic condition of the form  $|\dot{\lambda}/\lambda T|\ll1$, where $T$ is the time scale of the scattering process. Within this regime, higher order interactions should be suppressed. Although this might relax the strong coupling bound coming from the scattering amplitudes, it is not clear how time dependence would affect the other strong coupling scales, as treated in detail in \cite{Adshead:2014sga}.

\subsection{C. Validity of the effective single-field theory in the light of BICEP2}


In this section we study the relationship between the rate of change of the speed of sound and an adiabatic evolution, or in other words, how strong a turn can be without invalidating the single-field description. Particle production due to sudden turns has been previously studied (see e.g. \cite{Konieczka:2014zja} and references therein), and it constitutes an important consistency check for a valid single-field description. However, the situation has become much more exciting in the light of the new results of BICEP2 \cite{Ade:2014xna}, which pose an interesting challenge for effective single-field theories, as we will explain below. Let us first discuss the adiabatic condition in the context of integration of a heavy mode. The validity of the effective single-field theory is subject to the adiabatic condition \cite{Cespedes:2012hu}:
\begin{equation}
|\ddot {\cal F}_{\cal R}|\ll M_\text{eff}^2|{\cal F_R}|\ ,
\end{equation}
where ${\cal F_R}$ is the isocurvature fluctuation, associated to the heavy mode, which we integrate out to get an effective single-field description for the adiabatic curvature perturbation. $M_\text{eff}$ is the effective mass of the heavy field, determined by the turning rate in field space, the curvature of the scalar potential in the heavy direction, and the curvature of the field manifold (see e.g. \cite{Achucarro:2012sm}). The above condition can be recast in terms of background quantities as follows\footnote{We are disregarding a short transient at the start and end of the turn, where a different condition is satisfied.} \cite{Cespedes:2012hu}:
\begin{equation}
\left|\frac{d}{dt}\ln \(c_s^{-2}-1\)\right|\ll M_\text{eff}\ .
\label{adiabatic2}
\end{equation}
In a slow-roll regime, the conformal time is $\tau\simeq -{1}/{(aH_0)}$, and therefore $H_0\,dt=-d\tau/\tau$. Using this relation, we can rewrite the adiabatic condition \eref{adiabatic2} as follows:
\begin{eqnarray}
|s|&\ll&\frac{1}{2}\(1-c_s^2\)\frac{M_\text{eff}}{H_0}\ .
\label{adiabatic3}
\end{eqnarray}
Since in this paper we are focusing on the regime $|s|<1$, it is worth evaluating when the adiabatic condition \eref{adiabatic3} is automatically satisfied given the requirement of not-so-sharp turns $|s|<1$. One can see that
\be
\text{if}\quad c_s^2<1-\frac{2H_0}{M_\text{eff}}\quad ,\text{ then}\quad |s|<1\ \Longrightarrow\  \text {Adiabatic}\ .
\label{adiabatic4}
\ee
Given that in a valid EFT one should have $M_\text{eff}\gg H_0$, it is clear that the condition $|s|<1$ will ensure an adiabatic evolution. In terms of the effective mass, from \eref{adiabatic4} one can see that when the effective mass satisfies the lower bound
\be
M_\text{eff}\gtrsim \frac{2H_0}{|u|}\ ,
\label{lowerboundm}
\ee
the regime $|s|< 1$ automatically implies that we are in an effective single-field regime.\footnote{We stress that \eref{lowerboundm} is not an adiabatic condition, it is the condition under which smooth turns ($|s|< 1$) imply an adiabatic regime. Even if the lower bound \eref{lowerboundm} is violated, the condition \eref{adiabatic2} will still ensure adiabaticity.} Note that these considerations apply to any effectively single-field inflationary scenario in which a large hierarchy of masses and slow-roll are respected.

Now let us turn the discussion to the possibilities one has to achieve an effective single-field regime in the light of the new BICEP2 results. In this context, the main concern raised by their results is that a large tensor-to-scalar ratio sets the inflationary scale to a value close to the GUT scale, and therefore the energy gap in which the inflaton and the possible additional UV degrees of freedom must cohabit is not very large. Given this, having a large hierarchy of masses does not seem so easy.

Putting in some numbers, a naive interpretation of $r={\cal O}(0.1)$ would support $H_0\sim 10^{14}$ GeV \cite{Kaloper:2014zba}, leaving four orders of magnitude to the Planck scale. If there is new physics at the GUT scale (or above), then $|s|<1$ and $10^{-2}\lesssim |u|<1$ should be safely in the effectively single-field regime. Then, one could conclude that reductions in the speed of sound of a few percent are well motivated, and that the bound $|s|<1$ implies an adiabatic regime.

Summarizing, the new results by BICEP2, if confirmed, leave about five orders of magnitude in which the UV degrees of freedom and the inflaton must live together. Although the energy gap is not gigantic, one would expect the heavy physics energy scale to be at least a hundred times larger than the Hubble scale, and therefore the adiabatic condition is satisfied.


\subsection{D. Review of our search and further analyses}


In our previous paper \cite{Achucarro:2013cva}, we looked for correlated signatures in the primordial power spectrum and bispectrum due to a gaussian reduction in the speed of sound. We found several fits to the Planck CMB power spectrum data with an improvement\footnote{A similar result is obtained in the Standard Clock model \cite{Chen:2014joa}} $2<-\Delta \chi^2_\text{eff}<10$, and calculated the predicted correlated signals in the primordial bispectrum, whose shape turned out to be surprisingly similar to a set of primordial bispectrum templates tested against CMB bispectrum data by the Planck collaboration \cite[sec.\ 7.3.3]{Ade:2013ydc}.

Thanks to this similarity, we were able to qualitatively compare some of our predictions to some of their fits, finding a reasonable agreement \cite{Achucarro:2013cva}. But we also found interesting differences: (1) the analysis of \emph{localized} oscillations in the bispectrum performed by Planck only covers the region around the first acoustic peak, while our features are more significant around the second and third; (2) some of our best fits occur at values of $|\tau_0|$ corresponding to oscillatory frequencies which are slightly higher than those covered in Planck's analysis. Thus, an extended search for oscillatory features in the bispectrum data towards higher frequencies and smaller scales would help in confirming or falsifying our predictions. Although our fits are not very significant at the level of the CMB power spectrum, the mild agreement in the primordial bispectrum is more than encouraging, given that this prediction is solely based on a fit to the CMB power spectrum data, and that it comes from a well motivated and consistent theoretical framework.


\subsubsection{Review of main results and numerical consistency check}


The power spectrum features caused by a transient reduction in the speed of sound described by eq.\ \eqref{eq:gaussefolds}, parametrized by $B$, $\beta$ and $\tau_0$, are combined with the primordial spectrum of the \lcdm Planck baseline model described in \cite[sec.\ 2]{Ade:2013zuv}, parametrized by an amplitude $A_s$ and a spectral index $n_s$. The primordial perturbations evolve in a flat FLRW universe parametrized by the densities of baryonic and cold dark matter, $\Omega_\mathrm{b}$ and $\Omega_\mathrm{cdm}$, and the current expansion rate $H_0$. The damping due to reionization is parametrized by the optical depth $\tau_\mathrm{reio}$. Those 6 standard plus 3 feature parameters describe our cosmological model.

The features given by eq.\ \eqref{eq:deltappfourier} are calculated using a Fast Fourier Transform. The resulting CMB, calculated with the Boltzmann code \class \cite{Lesgourgues:2011re,Blas:2011rf}, is fitted to the ESA Planck mission temperature data of March 2013, using the likelihood provided by the experiment \cite{Ade:2013kta}, and the low-$\ell$ CMB polarization data of the WMAP experiment \cite{Bennett:2012zja}. In that fit, we use flat priors on the 6 cosmological parameters and on $B$, $\ln\beta$ and $\ln(-\tau_0)$. The bounds on the priors are those defined in \eqref{eq:bounds} and \eqref{eq:2bound3}, ignoring a priori the bound $|B|\gg\mathcal{O}(\epsilon,\eta)$. The posterior probability is then maximized over the prior bounds using Markov-Chain Monte-Carlo (MCMC) methods, making use of the MCMC sampler \montepython \cite{Audren:2012wb}.

\begin{figure}[t!]
\centering
\includegraphics[width=0.45\linewidth]{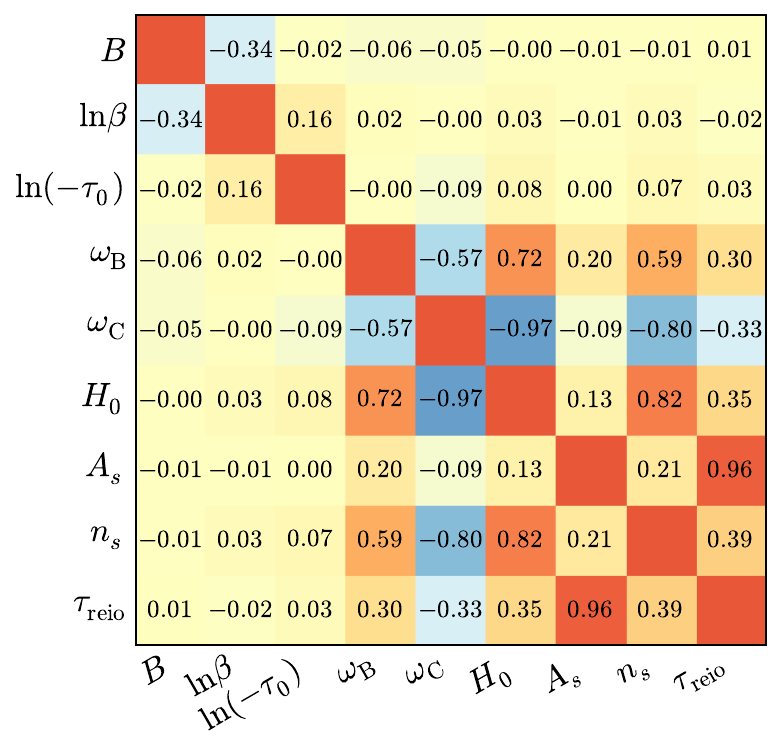}
\caption{\label{fig:corrs}
Correlation coefficients between the feature and the cosmological parameters for the mode \mode{B}. Notice the small correlations between the two sets of parameters, and the rather large negative correlation between $B$ and $\ln\beta$.}
\end{figure}

As is usual when fitting small features on top of a large data set, we found the likelihood (and hence the posterior) probability distribution to be multi-modal. As our features are small and affect only a fraction of the data set, we expect to find only mild degeneracies of the feature parameters with the cosmological parameters. Due to the mild character of the degeneracies (that we confirmed a posteriori, cf.\ fig.\ \ref{fig:corrs}), we expect the likelihood to show its multi-modal character only within the parameter subspace of the feature. Therefore, we start our search by mapping the multi-modal likelihood on this 3-dimensional subspace. When the position and extension of the modes were sufficiently well determined, we cropped unimodal regions and sampled them allowing now the cosmological parameters (and also the likelihood nuisance parameters) to vary. With this, we got the definitive posterior probability distribution functions for the different modes.

\begin{figure}[t!]
\includegraphics[width=0.75\linewidth]{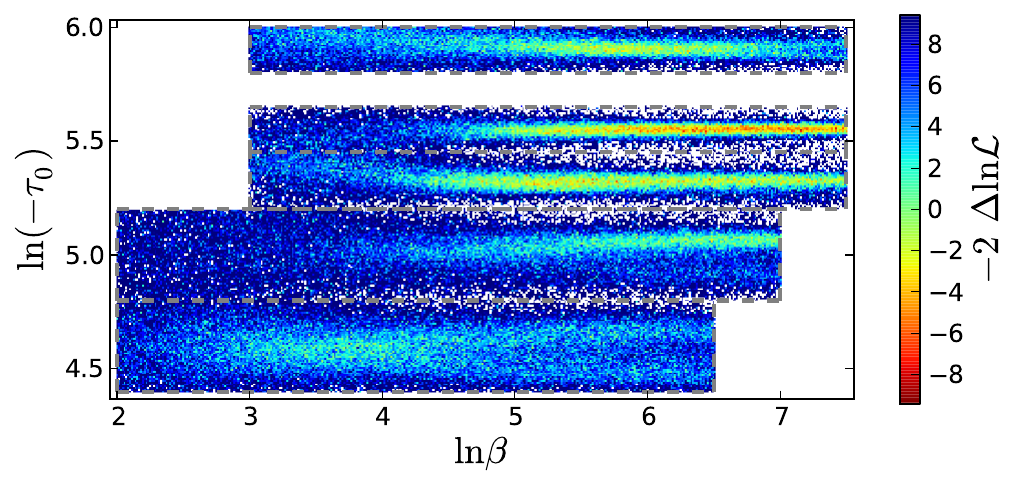}
\caption{Profile of $\Delta\chi^2_\text{eff}=-2\Delta\ln\mathcal{L}$ for the features in the CMB power spectrum in the $\left(\ln\beta,\,\ln(-\tau_0)\right)$ plane \cite{Achucarro:2013cva}.}
\label{fig:modes}
\end{figure}

Here we reproduce the results published in \cite{Achucarro:2013cva}, in a little more detail. The resulting profile likelihood can be seen in figure \ref{fig:modes}. There, one can identify five \emph{modes}, or defined regions of the parameter space where the likelihood is improved. The improvement is shown in $\Delta\chi^2_\text{eff}$, with $\chi^2_\text{eff} \equiv -2\ln\mathcal{L}$, and $\Delta$ meaning the difference with respect to the likelihood of the \lcdm Planck baseline model: $\chi^2_\text{eff}=9805.90$, using the data sets mentioned above.\footnote{See the parameter tables at \url{http://www.sciops.esa.int/wikiSI/planckpla/index.php?title=File:Grid_limit68.pdf&instance=Planck_Public_PLA}.} Regions with improvements of $-\Delta\chi^2_\text{eff} < 2$ have been discarded and are not shown in the plot.


\begin{table}
\newcommand{\btf}[1]{\ensuremath{(#1)}}
\newcommand{\btfint}[4]{$\btf{#1}\,#2\,\substack{+#3\\-#4}$}
\newcommand{\btfintsimple}[3]{$\btf{#1}\,#2\,\pm #3$}
\newcommand{\colspacing}{~~~}
\centering
\scalebox{1}{
\begin{tabular}{c@{\colspacing}c@{\colspacing}c@{\colspacing}c@{\colspacing}c@{\colspacing}c}
Mode & $-B\times 10^{2}$ & $\ln\beta$ & $\ln(-\tau_0)$ & $\Delta\chi^2_\text{eff}$ & $s_\text{max}$\\
\hline\hline
\mode{A} & \btfint{4.5}{3.7}{1.6}{3.0}
  & \btfint{5.7}{5.7}{0.9}{1.0}
  & \btfint{5.895}{5.910}{0.027}{0.035}
  & $-4.3$
  & $0.33$\\[1mm]
\hline
\mode{B} & \btfintsimple{4.2}{4.3}{2.0}
  & \btfint{6.3}{6.3}{1.2}{0.4}
  & \btfint{5.547}{5.550}{0.016}{0.015}
  & $-8.3$
  & $0.42$\\[1mm]
\hline
\mode{C} & \btfint{3.6}{3.1}{1.6}{1.9}
  & \btfint{6.5}{5.6}{1.9}{0.7}
  & \btfint{5.331}{5.327}{0.026}{0.034}
  & $-6.2$
  & $0.40$\\[1mm]
\hline
\mode{D} & \btf{4.4}
  & \btf{6.5}
  & \btf{5.06}
  & $-3.3$
  & $0.48$\\[1mm]
\hline
\mode{E}$^*$ & \btf{1.5}
  & \btf{4.0}
  & \btf{4.61}
  & $-2.2$
  & $0.05$
\end{tabular}}
\caption{\label{tab:paramranges}
CMB power spectrum best fits (in parentheses), $68\%$ c.l.\ intervals and effective $\Delta\chi^2$ at the best fit value for each of the modes. The prediction for the bispectrum for \mode{E} is not reliable (see \cite{Achucarro:2013cva}).}
\end{table}


\begin{figure}[t!]
\centering
\includegraphics[width=0.4\linewidth]{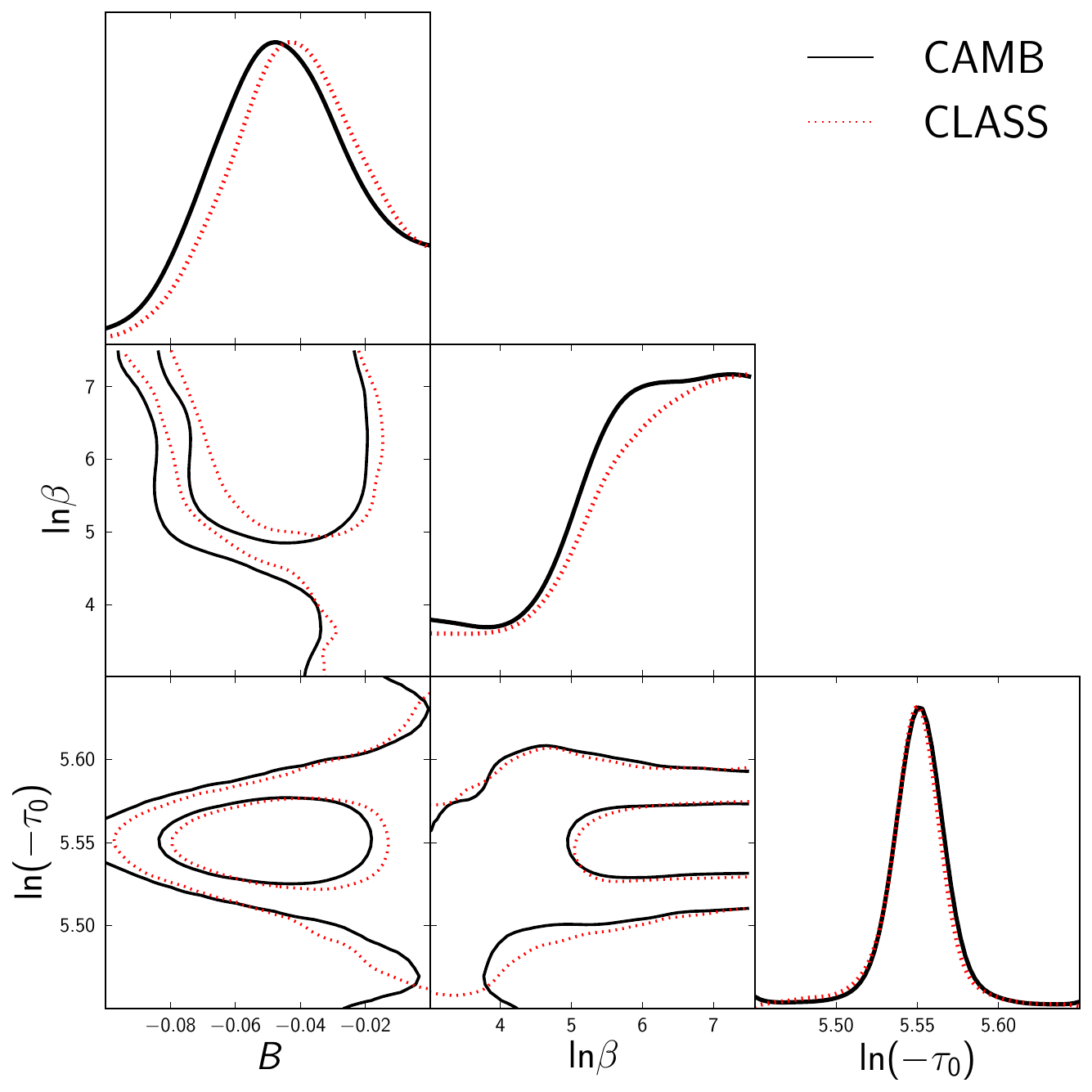}
\caption{\label{fig:camb_vs_class}
\camb\unskip$+$\cosmomc vs.\ \class\unskip$+$\montepython consistency check: 1D and 2D marginalized posterior distributions of the sound speed reduction parameters for the mode \mode{B}.
}
\end{figure}


\setlength\tabcolsep{1pt}
\begin{table*}[t!]
\footnotesize
\centering
\begin{tabular}{|l|c|c|c|}
\hline
\multicolumn{4}{|c|}{\textit{\textbf{Planck}}+\bf WP}\\
\hline
\hline
Parameter                                            &  CAMB                             & CLASS   &  Baseline \cite{Ade:2013zuv}\\ 
\hline
$100\Omega_b h^2$                           & $2.208\pm0.027$                   & $2.214\pm0.029$  & $2.205\pm0.028$\\
$\Omega_c h^2$                              & $0.1204\pm0.0026$                 & $0.1203\pm0.0027$ & $0.1199\pm0.0027$\\
$\tau_\text{reio}$                                      & $0.089\pm0.013$                   & $0.090\pm0.013$ & $0.089\substack{+0.012\\-0.014}$\\
$H_0 $
& $67.16\pm1.14$                    & $67.29\pm1.21$ & $67.3\pm1.2$\\
$n_s$                                       & $0.9600\pm0.0070$                 & $0.9598\pm0.0074$ &
$0.9603\pm0.0073$ \\
$\ln(10^{10} A_s)$                         & $3.090\pm0.023$                   & $3.088\pm0.024$ & $3.089\substack{+0.024\\-0.027}$\\
\hline
$B$                                         & $-0.045^{+0.045}_{-0.034}$ (95$\%$c.l.)      & $-0.041^{+0.041}_{-0.031}$ (95$\%$c.l.) & --- \\
$\ln\beta$                                 & $6.00^{+1.50}_{-3.00}$  (95$\%$c.l.)      & $6.06^{+1.44}_{-2.18}$ (95$\%$c.l.) & --- \\
$\ln(-\tau_0)$                             & $5.55\pm0.06$  (95$\%$c.l.)                    & $5.55\pm0.05$ (95$\%$c.l.)& --- \\
\hline
\hline
$\chi^2_\text{bf}$                        &  9797.25                         & 9797.58 & 9805.90 \\
\hline
\end{tabular}
\caption{
\camb\unskip$+$\cosmomc vs.\ \class\unskip$+$\montepython consistency check: mean values and $68\%$ (or $95\%$ where indicated) confidence intervals for the primary $\Lambda$CDM parameters and the additional sound speed reduction parameters for the mode \mode{B}. We also show the parameter ranges found by the Planck collaboration \cite{Ade:2013zuv} for a featureless model.}
\label{Tab:camb_vs_class}
\end{table*}


As can be seen in figure \ref{fig:modes}, the modes are well-isolated narrow bands of $\ln(-\tau_0)$, i.e. frequency of oscillation of the primordial spectrum feature. For each of the modes showed in the figure, the relevant parameter data is given in table \ref{tab:paramranges}: the numbers in parentheses are the best fit values, and the parameter ranges, when given, are $68\%$ c.l.\ regions.

The upper limit for $\ln\beta$ in the modes \mode{B} and \mode{C} is imposed by the prior, as we will explain below. For the modes \mode{D} and \mode{E}, no parameter ranges are given, due to their low significance and non-gaussian character; only the respective best fits are shown.

As expected, we find only small degeneracies\footnote{The correlation matrix is defined as $\rho_{ij}\equiv C_{ij}/\sqrt{C_{ii}\cdot C_{jj}}$, where $C_{ij}$ are the covariance matrix elements corresponding to the parameters with indices $i$ and $j$.} ($|\rho|\le0.15$) between the feature parameters and the \lcdm parameters for modes \mode{A}, \mode{B} and \mode{C}. Consequently, the best fits and $68\%$c.l.\ intervals of the \lcdm parameters reproduce quite accurately those of Planck, cf. table \ref{Tab:camb_vs_class}. The correlation matrix for the mode \mode{B} is shown in fig.\ \ref{fig:corrs}. For the less significant modes \mode{D} and \mode{E}, some of the correlations grow up to $|\rho|\le0.30$. This is expected, since for lower $\ln(-\tau_0)$ the frequency of the fits drops, getting closer to the frequency of the acoustic oscillations.

\begin{figure}[t!]
\centering
\includegraphics[width=0.74\linewidth]{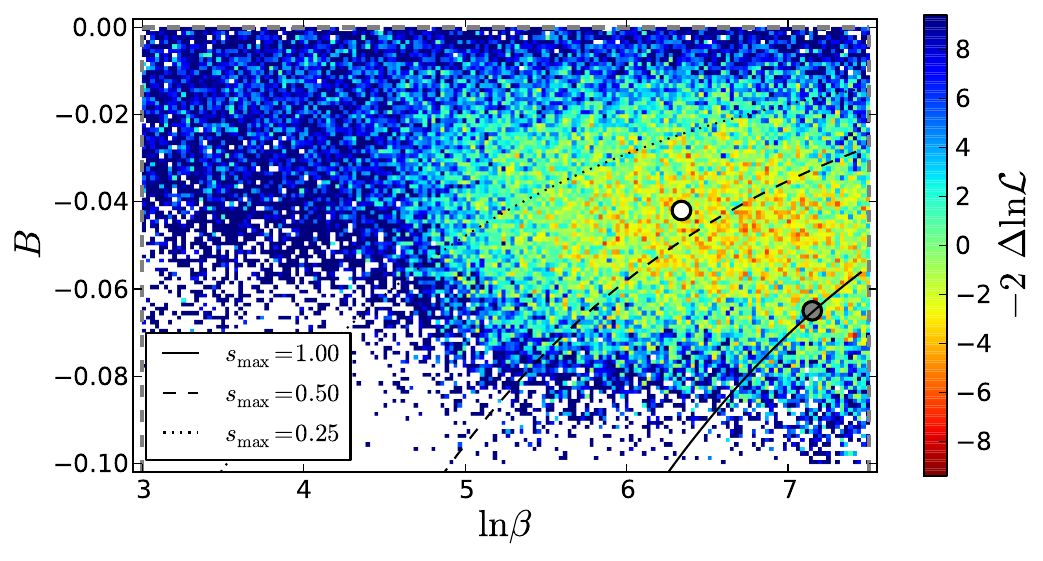}
\caption{\label{fig:degBbeta_modeB}
Profile of $\Delta\chi^2_\text{eff}=-2\Delta\ln\mathcal{L}$ for the mode \mode{B} in the $(\ln\beta,\,B)$ plane, showing the $\rho=-0.34$ degeneracy between those two parameters. Some lines of $s_\text{max}=\text{const}$ are shown. Notice how the mode extends beyond the $s=1$ prior limit.}
\end{figure}


\begin{figure*}[t!]
\centering
\subfigure{\includegraphics[width=0.44\linewidth]{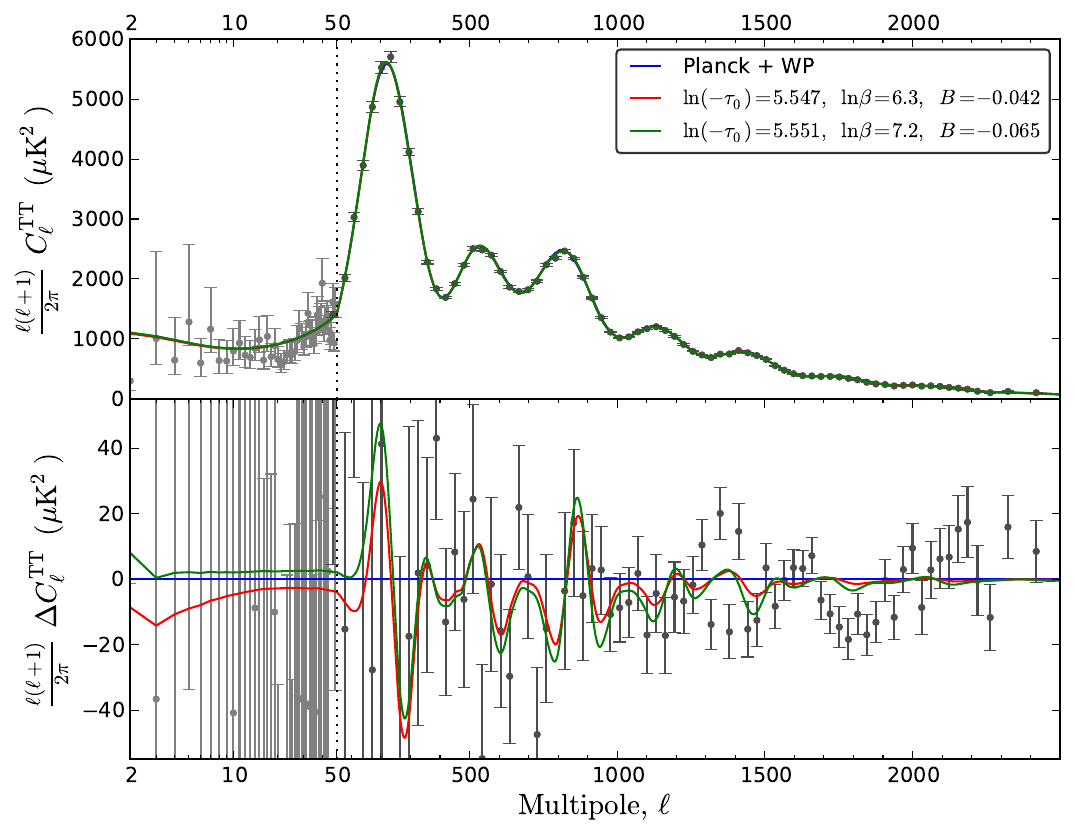}}
\subfigure{\includegraphics[width=0.44\linewidth]{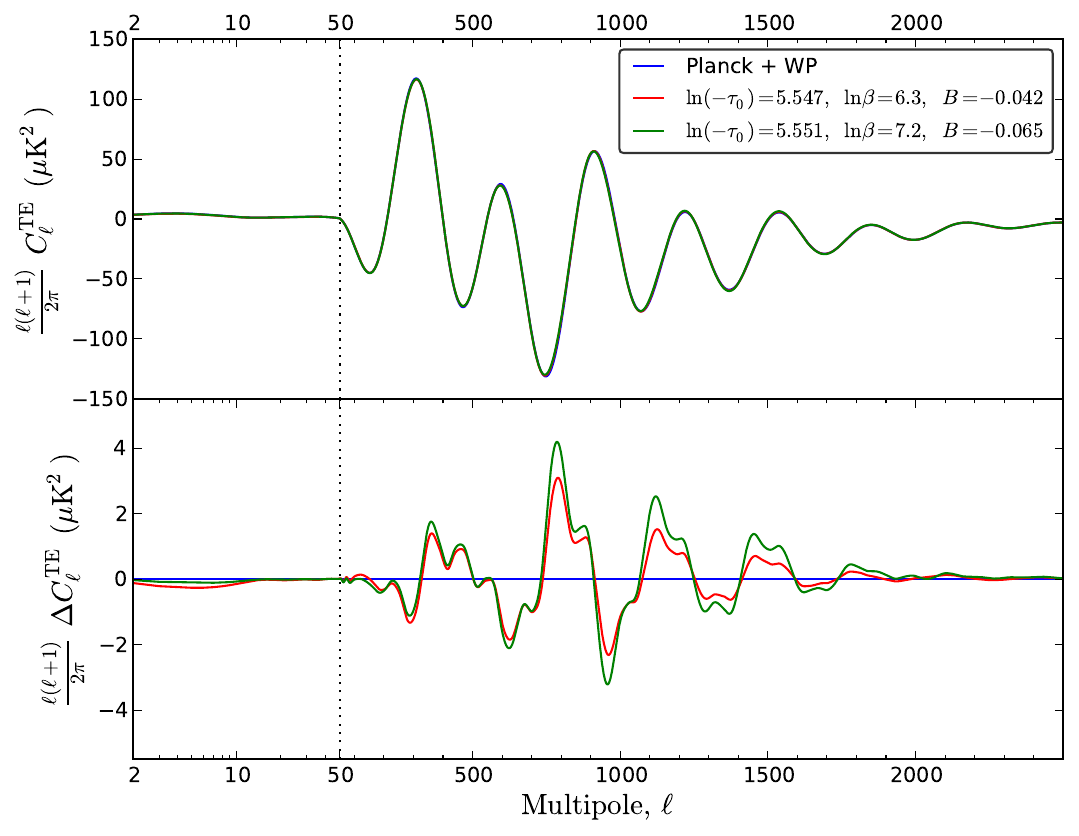}}
\subfigure{\includegraphics[width=0.44\linewidth]{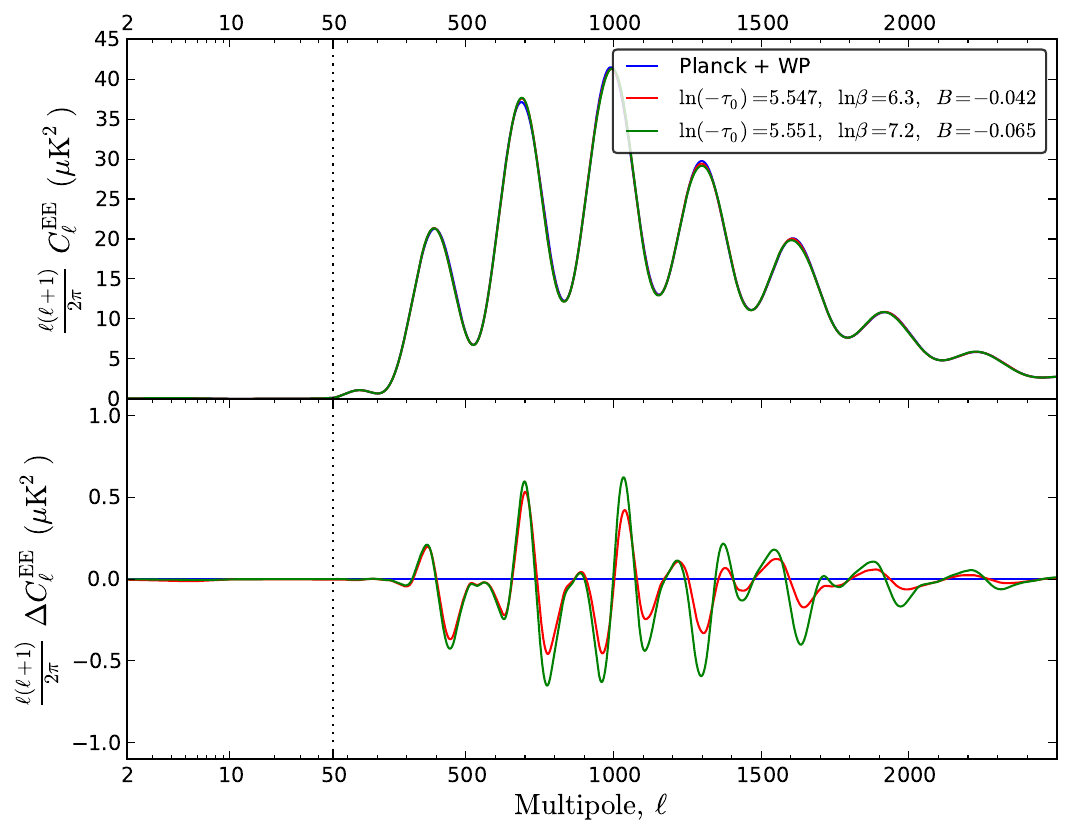}}
\caption{\label{fig:comp2bf}
Comparison of the two fits indicated in figure \ref{fig:degBbeta_modeB} with a white circle (red, dashed line) and a gray circle (green, dotted line), in the TT, TE and EE CMB power spectra.}
\end{figure*}


In order to make our results from \class\unskip$+$\montepython more reliable, we cross-checked them with an independent Einstein-Boltzmann solver and a different MCMC sampler, namely \camb \cite{Lewis:1999bs} and \cosmomc \cite{Lewis:2002ah}. As an example, in fig.\ \ref{fig:camb_vs_class} and tab.\ \ref{Tab:camb_vs_class} we explicitly show this comparison for the most significant mode \mode{B} by varying both the primary $\Lambda$CDM parameters and the additional sound speed reduction parameters. We find excellent agreement between these two results.


\subsubsection{Degeneracies in the modes and polarization}


The CMB temperature data is not able to restrict the maximum value of $\ln\beta$, as one can see in figure \ref{fig:modes} and in the 1D marginalized likelihood of $\ln\beta$ in figure \ref{fig:camb_vs_class} (middle-right panel). After some value of it, the likelihood reaches a plateau with constant $\ln(-\tau_0)$ and increasing $\ln\beta$. As for the amplitude $B$, it is correlated with $\ln\beta$ with correlation coefficient of order $\rho\sim-0.3$ (cf.\ fig.\ \ref{fig:corrs}) -- we find that the best fit for $|B|$ increases along increasing $\ln\beta$ in each of these plateaus.

The reason for the data not being able to restrict $\ln\beta$ and for this degeneracy is quite well explained by figure \ref{fig:degBbeta_modeB} and figure \ref{fig:comp2bf}. In the last one, we have plotted the CMB temperature and E-mode polarization spectra of the best fit of the mode \mode{B} (white circle in figure \ref{fig:degBbeta_modeB}), together with a similar fit (grey circle in figure \ref{fig:degBbeta_modeB}) that improves $\Delta\chi^2_\text{eff}$ marginally and saturates the $s=1$ bound. Along the direction of simultaneous increase of $\ln\beta$ and $|B|$, the feature in the primordial spectrum broadens towards smaller scales, while the amplitude of the tail on the larger scales remains almost constant. Since at smaller scales much of the primordial signal is suppressed by diffusion damping in the CMB, no significance is gained along the degeneracy direction, causing a plateau in $\Delta\chi^2_\text{eff}$.

\begin{figure}[b!]
\centering
\includegraphics[width=0.49\textwidth]{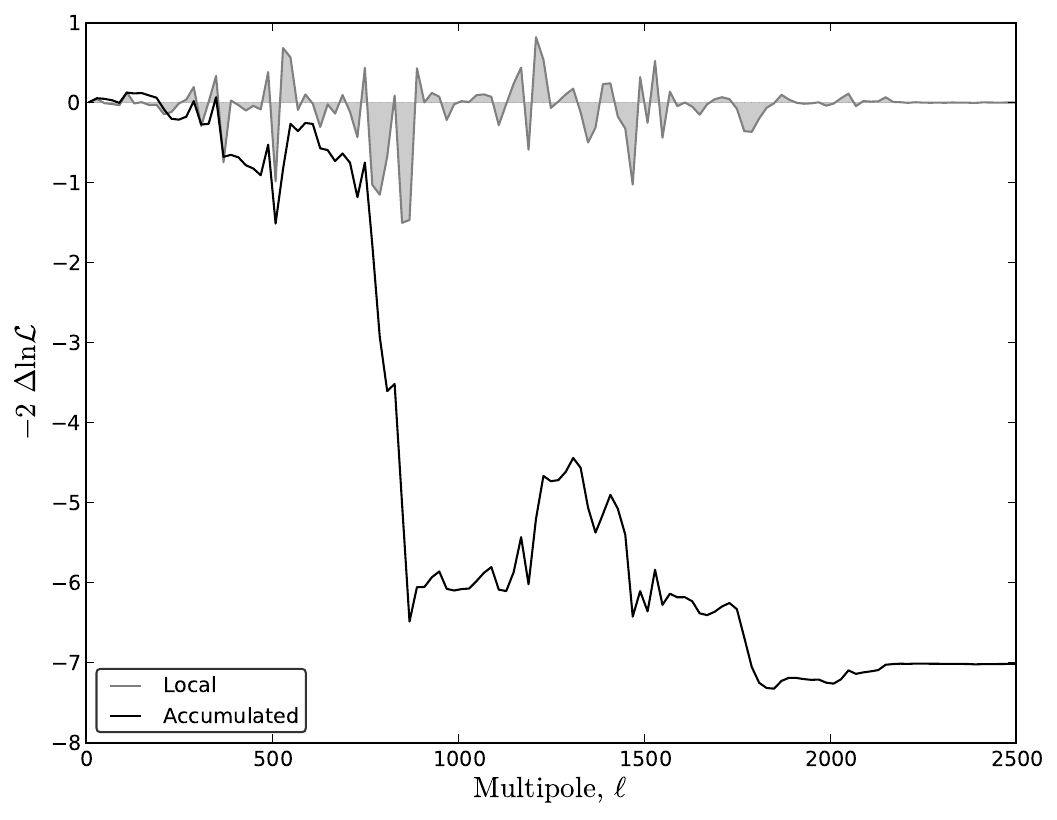}
\caption{\label{fig:chisq_l_B}
Gain in the likelihood of the best fit of mode \mode{B} along the multipoles. The gray area shows the local difference in each bin, and the black line shows the accumulated difference for increasing multipoles.}
\end{figure}

Photon diffusion at the last scattering surface has the effect of polarizing the CMB signal through Thomson scattering, so at smaller scales the polarization spectrum will contain information about the primordial spectrum, complementary to that of the temperature spectrum. Therefore, the difference at small scales between two fits in the same plateau (for example the red and the green spectra in figure \ref{fig:comp2bf}) is larger in the polarization spectra (TE and EE). This suggests that the Planck polarization data, expected to be released along 2014, may be able to set stringer bounds on the maximum value of $\ln\beta$.


\subsubsection{Local improvement at different angular scales: $\Delta\chi^2(\ell)$}


Given a fit to the CMB power spectrum of some feature model, it is interesting to know in which ranges of multipoles the feature describes the data better than the baseline \lcdm model. This kind of \emph{local improvement} can only be calculated approximately, since the temperature data points at different multipoles are in general correlated. Nevertheless, even a qualitative analysis can shed some light on where the feature fits better the data than the baseline model.
 
We have studied the local improvements along the multipoles of the four relevant fits, modes \mode{A} to \mode{D} (we show the result for mode \mode{B} in figure \ref{fig:chisq_l_B}). To do that, we have binned the multipoles with $\Delta\ell=20$ and substituted pieces of the best fit for each mode into the best fit of the \lcdm baseline model. For the sake of simplicity, we use for this analysis the preliminary fits found by keeping the cosmological and nuisance parameters fixed to their best fit values (hence the small difference in the total $\Delta\chi^2_\text{eff}$ between fig.\ \ref{fig:chisq_l_B} and tab.\ \ref{tab:paramranges}).

The results show that mode \mode{A} gains its significance mostly in the first and third peak and loses some of it in the second; mode \mode{B} (see fig.\ \ref{fig:chisq_l_B}) and \mode{C} gain most of their significance in the third peak, lose some of it in the fourth peak and improve a little again in the fifth and sixth. The mode \mode{D} does not fit well the first and second peaks, gains most of its significance in the third peak, and some more in the fifth and sixth peaks.


\section{IV. Conclusions and outlook}

A detailed understanding of the origin and detectability of transient features in the primordial (and observed) correlation functions is now more important than it was before the BICEP2 results \cite{Ade:2014xna}. A large transplanckian field excursion should detect any features present in the scalar potential as well as changes in the dispersion relation of the adiabatic mode, if they are there, and arguably there were hints of both in the Planck data \cite{Ade:2013uln,Ade:2013ydc}. At the same time, a high inflationary scale leaves less room for mass hierarchies in the UV completion, that would be needed to justify the single-field effective low energy description. This is a problem for very sharp features, as they tend to excite any higher frequency modes coupled to the inflaton. We have argued that the regime of moderately sharp features is particularly interesting. Most likely these cannot be detected in any particular dataset and have to be searched for in {\it correlations} between different data sets.

 In this regime, the effect of a transient reduction in the speed of sound can be calculated with the simple SRFT approximation \cite{Achucarro:2012fd}, in which the correlations between power spectrum and bispectrum are manifest. We emphasize that the simple expressions \eref{eq:deltappfourier} and \eref{bispectrumana} hold provided ${\cal O}(\epsilon, \eta)\ll \text{max}\(|1-c_s^{-2}|,\, |\dot c_s/(Hc_s)|\)\ll 1$ and $c_s=1$ before and after the feature.

In this work we have presented an alternative way to calculate both the power spectrum and bispectrum, by consistently applying an approximation for moderately sharp features, both to the GSR power spectrum (eq. \eref{eq:gsr3}) and
to the in-in calculation of the bispectrum (eq. \eref{completebi}). Within this regime, we have extended existing GSR calculations of the power spectrum to less sharp and arbitrary shapes of the speed of sound, and found excellent agreement with the SRFT approximation in the regime where both methods apply.

Given that the regimes of validity of the two methods are not entirely coincident, we are now equipped with a robust machinery that will allow us to describe features in the speed of sound for a broader region of the parameter space. Broad features can be calculated with the SRFT approach, while sharp features can be calculated using GSR for the power spectrum (eq. (\ref{eq:gsr3})) and the in-in approach for the bispectrum (eq. (\ref{completebi})).

In a previous paper \cite{Achucarro:2013cva} we performed a search for such correlated features assuming moderately sharp, mild reductions in the speed of sound of the adiabatic mode during uninterrupted slow-roll inflation. We reported several fits to the Planck CMB temperature spectrum data and predicted the correlated signatures in the complete primordial bispectrum. We qualitatively compared with the bispectrum search by Planck when possible and found reasonable agreement.
We have performed additional tests to the results of our search in \cite{Achucarro:2013cva}. Namely, we have repeated it using independent codes and found practically equal results; we have studied more explicitly the small degeneracies among the cosmological and feature parameters, and proposed the CMB TE and EE polarization spectra as a way to break degeneracies among the latter; and finally we have investigated at which multipoles each of our fits describe the CMB temperature data better than the baseline \lcdm model.

The ability to make predictions in a wider region of the parameter space of features is of particular relevance, since new data sets may allow us to explore it. Besides, since different experiments generally have different foregrounds and systematics, a joint analysis could reduce the contamination of the primordial signal on the overlapping scales. In particular, we plan to extend our search to large scale structure surveys \cite{features3}.




\subsection{Acknowledgments}
We are grateful to Dario Cannone and Gonzalo Palma for discussions. This work was
partially supported by the Netherlands Foundation for Fundamental
Research on Matter F.O.M., the DFG Graduiertenkolleg ``Particle
Physics at the Energy Frontier of New Phenomena'', a Leiden Huygens
Fellowship, and the Netherlands Organization for Scientific Research
(NWO/OCW) under the Gravitation Program. Also by the Spanish Ministry
of Science and Technology Grant FPA2012-34456, the Spanish
Consolider-Ingenio 2010 Programme CPAN (CSD2007-00042) and the Basque
Government Grant IT559-10.



 \newcommand{\noop}[1]{}

\end{document}